\RequirePackage[l2tabu,orthodox]{nag}
\RequirePackage{fix-cm}
\documentclass[12pt,a4paper]{article}

\usepackage{fixltx2e}

\usepackage[font=small]{caption}
\usepackage[text={450pt,650pt},headheight={14.5pt},centering]{geometry}

\usepackage{lmodern}

\usepackage{graphicx}
\usepackage{booktabs}
\usepackage{subfig}
\usepackage{tikz}
\usetikzlibrary{calc,decorations,decorations.pathmorphing}

\tikzstyle dynkin node=[very thick,shape=circle,draw,inner sep=0pt,minimum size=5mm]
\tikzstyle dynkin line=[very thick]
\tikzstyle inverse line=[gray,very thick,densely dotted]
\tikzstyle red phase=[red,decoration={snake,amplitude=0.1mm,segment length=1.6mm},decorate]
\tikzstyle blue phase=[blue,decoration={snake,amplitude=0.1mm,segment length=0.9mm},decorate]

\usepackage[numbers,merge]{natbib}
\usepackage{collnat}

\usepackage{amsmath,amssymb}
\usepackage{mathtools}
\usepackage{bbm}
\usepackage{slashed}
\usepackage{braket}

\usepackage{dsfont}

\DeclareMathAlphabet{\mathsfit}{\encodingdefault}{\sfdefault}{m}{sl}

\usepackage{hyperref}

\usepackage{xspace}

\numberwithin{equation}{section}

\makeatletter
 \let\old@startsection=\@startsection
 \let\oldl@section=\l@section
 \renewcommand{\@startsection}[6]{\old@startsection{#1}{#2}{#3}{#4}{#5}{#6\mathversion{bold}}}
 \renewcommand{\l@section}[2]{\oldl@section{\mathversion{bold}#1}{#2}}
\makeatother

\def\Xint#1{\mathchoice
  {\XXint\displaystyle\textstyle{#1}}%
  {\XXint\textstyle\scriptstyle{#1}}%
  {\XXint\scriptstyle\scriptscriptstyle{#1}}%
  {\XXint\scriptscriptstyle\scriptscriptstyle{#1}}%
  \!\int}
\def\XXint#1#2#3{{\setbox0=\hbox{$#1{#2#3}{\int}$}
    \vcenter{\hbox{$#2#3$}}\kern-.5\wd0}}
\def\pint{\;\Xint-}

\newcommand{\AdS}{\textup{AdS}}
\newcommand{\CFT}{\textup{CFT}}
\newcommand{\Sphere}{\textup{S}}
\newcommand{\Torus}{\textup{T}}

\newcommand{\Smat}{\mathcal{S}}

\newcommand{\alg}[1]{\mathfrak{#1}}
\newcommand{\grp}[1]{\mathrm{#1}}

\newcommand{\algD}[1]{\alg{d}(2,1;#1)}
\newcommand{\grpD}[1]{\grp{D}(2,1;#1)}

\newcommand{\algSU}{\alg{su}}
\newcommand{\grpSU}{\grp{SU}}

\newcommand{\gen}[1]{\mathfrak{#1}}

\newcommand{\genQ}{\gen{Q}}
\newcommand{\genS}{\gen{S}}
\newcommand{\genH}{\gen{H}}

\newcommand{\genP}{\gen{P}}
\newcommand{\genK}{\gen{P}^\dag}

\newcommand{\superN}{\mathcal{N}}

\newcommand{\ie}{\textit{i.e.}\xspace}
\newcommand{\eg}{\textit{e.g.}\xspace}

\newcommand{\fixedspaceL}[2]{\mathrlap{#2}\phantom{#1}}
\newcommand{\fixedspaceR}[2]{\phantom{#1}\mathllap{#2}}

\newcommand{\smallL}{\scriptscriptstyle\textit{L}}
\newcommand{\smallR}{\scriptscriptstyle\textit{R}}

\newcommand{\I}{\text{I}}
\newcommand{\II}{\text{II}}

\begin{document}

\thispagestyle{empty}

\begin{flushright}\footnotesize\ttfamily
ITP-UU-12/48\\
SPIN-12/45
\end{flushright}
\vspace{5em}

\begin{center}
\textbf{\Large\mathversion{bold} All-loop Bethe ansatz equations for $\AdS_3/\CFT_2$}

\vspace{2em}

\textrm{\large Riccardo Borsato, Olof Ohlsson Sax and Alessandro Sfondrini} 

\vspace{2em}

\textit{Institute for Theoretical Physics and Spinoza Institute,\\ Utrecht University, 3508 TD Utrecht, The Netherlands}

\vspace{1em}

\texttt{R.Borsato@uu.nl, O.E.OlssonSax@uu.nl, A.Sfondrini@uu.nl}


\end{center}

\vspace{6em}

\begin{abstract}\noindent
Using the S-matrix for the $\algD{\alpha}^2$ symmetric spin-chain of $\AdS_3/\CFT_2$, we propose a new set of all-loop Bethe equations for the system. These equations differ from the ones previously found in the literature by the choice of relative grading between the two copies of the $\algD{\alpha}$ superalgebra, and involve four undetermined scalar factors that play the role of dressing phases. Imposing crossing symmetry and comparing with the near-BMN form of the S-matrix found in the literature, we find several novel features. In particular, the scalar factors must differ from the Beisert-Eden-Staudacher phase, and should couple nodes of different masses to each other.
In the semiclassical limit the phases are given by a suitable generalization of Arutyunov-Frolov-Staudacher phase.
\end{abstract}

\newpage

\section{Introduction}
\label{sec:introduction}
The $\AdS/\CFT$ correspondence is a striking and thought-inspiring conjecture. The strongly coupled regime of gravity theories, such as superstring theories, on $\AdS_{d+1}\times X$ with $X$ a compact space should be dual to a weakly coupled, $d$-dimensional, conformal field theory (CFT) on the boundary of $\AdS_{d+1}$, and vice-versa \cite{Maldacena:1997re,Witten:1998qj,Gubser:1998bc}. This can allow us to deepen our understanding of gravity and quantum field theory.

Another remarkable aspect of the correspondence is that, in the 't Hooft limit, integrable structures may arise on both sides of the correspondence. The first and most studied example of integrability in this context is $\AdS_5/\CFT_4$ that relates superstrings on a Ramond-Ramond (RR) $\AdS_5\times \Sphere^5$ background to $\mathcal{N}=4$ supersymmetric Yang-Mills theory (SYM); there, one is in principle able to compute the exact spectrum of any string state by integrability, and match it with perturbative calculations in either theory, see~\cite{Arutyunov:2009ga,Beisert:2010jr} for a review.

Here we will focus on another instance of the duality where integrability seems to be playing a role, that is when the superstring RR background is $\AdS_3\times \Sphere^3\times \Sphere^3\times \Sphere^1$. This preserves 16 supercharges provided that the radii of the warped spaces satisfy the triangle equality
\begin{equation}
  \frac{1}{R_{\AdS}^2} = \frac{1}{R_1^2} +  \frac{1}{R_3^2} ,
\end{equation}
where $1$ and $3$ denote the two three-spheres. We are therefore dealing with a family of backgrounds  that can be labeled by
\begin{equation}
  \alpha = \frac{R_{\AdS}^2}{R_1^2} = 1 - \frac{R_{\AdS}^2}{R_3^2} ,
\end{equation}
with $0<\alpha<1$, each corresponding to a two-dimensional CFT with large $\mathcal{N}=4$ superconformal symmetry~\cite{deBoer:1999rh,Gukov:2004ym,Berg:2006ng}.
The Green-Schwarz action for the superstrings can be rewritten as a supersymmetric coset model on~\cite{Babichenko:2009dk}
\begin{equation*}
  \frac{\grpD{\alpha}\times\grpD{\alpha}}{\grpSU(1,1)\times \grpSU(2)\times \grpSU(2)},
\end{equation*}
plus an additional decoupled massless scalar from the $\Sphere^1$ in the background. The advantage of the coset description is that it makes classical integrability manifest, from which finite gap equations can be written down~\cite{Babichenko:2009dk,Zarembo:2010yz}. These describe the semiclassical string spectrum, and are a limit of the all-loop Bethe ansatz (BA) equations that describe the full asymptotic spectrum. In~\cite{Babichenko:2009dk} knowledge of the finite gap equations was used to reverse-engineer the all-loop BA, that was further analyzed in~\cite{OhlssonSax:2011ms} where an alternating symmetric $\alg{d}(2,1;\alpha)^2$ spin-chain description was derived.

The spectrum described by the finite gap equations is missing two massless modes. One mode corresponds to excitations on $\Sphere^1$, and the other to a mode shared by the two spheres that is not present in the coset model since the Virasoro constraints are overimposed there. These modes can be put back by hand at the classical level \cite{Babichenko:2009dk,Sundin:2012gc}, but it is not yet clear how to do that at the quantum level. See~\cite{Sax:2012jv} for a discussion about massless modes in the weakly coupled spin-chain.

Here we want to take a different route: in~\cite{Borsato:2012ud}, the all-loop S-matrix was bootstrapped out of the symmetries of the theory.\footnote{%
  A different all-loop S-matrix was recently proposed in~\cite{Ahn:2012hw}. That S-matrix reproduces the Bethe equations of~\cite{Babichenko:2009dk,OhlssonSax:2011ms}, but as discussed in~\cite{Borsato:2012ud}, the underlying spin-chain does not have a well-defined notion of length. For further discussion on this point, see appendices~D and~F of~\cite{Borsato:2012ud}.%
} %
 Building on that, it is a relatively straightforward task to write down the resulting Bethe ansatz equations. However, we will find some unexpected features. In particular the resulting equations are written in a different, and seemingly inequivalent, grading of $\alg{d}(2,1;\alpha)^2$ compared to the one used in \cite{Babichenko:2009dk,OhlssonSax:2011ms}. Furthermore, they feature new couplings between particles of different mass. We also find that the scalar factors that appear in the BA, while not being simply related to the Beisert-Eden-Staudacher dressing phase of $\AdS_5/\CFT_4$ \cite{Beisert:2006ez} and $\AdS_4/\CFT_3$ \cite{Ahn:2008aa}, are expressible in terms of the one of Arutyunov, Frolov and Staudacher \cite{Arutyunov:2004vx} in the semiclassical limit. 

The plan of the paper is as follows. In section~\ref{sec:S-matrix-recap} we will briefly recall the form of the all-loop S-matrix of~\cite{Borsato:2012ud}; this involves some undetermined scalar factors, that are however constrained by a set of crossing equations, which we also present. In section~\ref{sec:diagonalisation} it is shown how the \emph{nesting procedure} can be used to diagonalise the S-matrix, in a way similar to \cite{Beisert:2005wm,Beisert:2005tm,Beisert:2005fw,deLeeuw:2007uf}. In section~\ref{sec:BA-equations} the resulting Bethe equations are presented, their fermionic dualization is constructed and the constraints on the scalar factors are discussed. Section~\ref{sec:comparison} compares our results with other known results in the finite gap \cite{Babichenko:2009dk,Zarembo:2010yz} and near-BMN \cite{Rughoonauth:2012qd} limits. In doing this, we find what appears to be a contradiction between the two semiclassical results.
 
\section{The S-matrix}
\label{sec:S-matrix-recap}
The S-matrix for the $\alg{d}(2,1;\alpha)^2$ symmetric alternating spin-chain was derived in~\cite{Borsato:2012ud} by imposing invariance under the residual symmetry algebra preserved by the spin-chain ground-state, which is a centrally extended $\algSU(1|1)^2$ algebra. Here we recap the results that we need in order to construct the Bethe ansatz.

The fundamental excitations of the spin-chain are doublets of one boson and one fermion, that can have mass $s$ equal to either $\alpha$ or $1-\alpha$, and left or right chirality. We will denote the bosons by $\phi^i_p$ and the fermions by $\psi^i_p$, where $p$ is the momentum of the excitation and the index $i$ labels the remaining flavors. Table~\ref{tab:excitations} summarizes our notation.

\begin{table}
  \centering
  \begin{tabular}{cll}
    \toprule
    $s$& Left & Right \\
    \midrule
    $\alpha$    & $\phi^{1},\,\psi^{1}$ & $\phi^{\bar{1}},\,\psi^{\bar{1}}$ \\
    $1-\alpha$ & $\phi^{3},\,\psi^{3}$ & $\phi^{\bar{3}},\,\psi^{\bar{3}}$ \\
    \bottomrule
  \end{tabular}
  \caption{%
    Fundamental excitations of the $\alg{d}(2,1;\alpha)^2$ symmetric alternating spin-chain divided by chirality (columns) and mass $s$ (rows).
  }
  \label{tab:excitations}
\end{table}

The matrix elements are naturally expressed in terms of the Zhukovski variables $x^\pm_p$, which satisfy
\begin{equation}
\label{eq:zhukovski}
  \frac{x_p^+}{x_p^-} = e^{ip}, \qquad
  \left(x_p^+ + \frac{1}{x_p^+}\right) - \left(x_p^- + \frac{1}{x_p^-}\right) = \frac{2is}{h},
\end{equation}
where $h>0$ is the coupling constant. We will generally denote the Zhukovski variables as  $x^\pm_p$ when $s=\alpha$ and as  $z^\pm_p$ when $s=1-\alpha$. As discussed in~\cite{Borsato:2012ud} the S-matrix is reflectionless and is naturally expressed in terms of the blocks appearing in the different sectors. 

\subsection{The \texorpdfstring{$11$, $\bar{1}\bar{1}$, $33$ and $\bar{3}\bar{3}$}{11, 1b1b, 33 and 3b3b} sectors}
In all these sectors we are scattering excitations having the same flavors. Then the S-matrix acts as
\begin{equation}
  \begin{aligned}
    \Smat \ket{\fixedspaceL{\psi^i_p \psi^i_q}{\phi^i_p \phi^i_q}} 
    &= \fixedspaceR{D^{ii}_{pq}}{A^{ii}_{pq}} \ket{\fixedspaceL{\psi^i_p \psi^i_q}{\phi^i_q \phi^i_p}} , \qquad &
    \Smat \ket{\fixedspaceL{\psi^i_p \psi^i_q}{\phi^i_p \psi^i_q}} 
    &= \fixedspaceR{D^{ii}_{pq}}{B^{ii}_{pq}} \ket{\fixedspaceL{\psi^i_p \psi^i_q}{\psi^i_q \phi^i_p}} + C^{ii}_{pq} \ket{\fixedspaceL{\psi^i_p \psi^i_q}{\phi^i_q \psi^i_p}}, \\
    \Smat \ket{\fixedspaceL{\psi^i_p \psi_q}{\psi^i_p \psi^i_q}} &= \fixedspaceR{D^{ii}_{pq}}{F^{ii}_{pq}} \ket{\fixedspaceL{\psi^i_p \psi^i_q}{\psi^i_q \psi^i_p}} , \qquad &
    \Smat \ket{\fixedspaceL{\psi^i_p \psi^i_q}{\psi^i_p \phi^i_q}} 
    &= \fixedspaceR{D^{ii}_{pq}}{D^{ii}_{pq}} \ket{\fixedspaceL{\psi^i_p \psi^i_q}{\phi^i_q \psi^i_p}} + E^{ii}_{pq} \ket{\fixedspaceL{\psi^i_p \psi^i_q}{\psi^i_q \phi^i_p}}, \\
  \end{aligned}
\end{equation}
where $i=1,\bar{1},3,\bar{3}$. When $i=1,\bar{1}$, the S-matrix elements assume the form
\begin{equation}
  \begin{aligned}
  \label{eq:Sii-elements}
    A^{ii}_{pq} &= + (S^{ii}_{pq})^{-1} \, \frac{x_q^+ - x_p^-}{x_q^- - x_p^+} , \ &
    B^{ii}_{pq} &= (S^{ii}_{pq})^{-1}  \, \frac{x_q^+ - x_p^+}{x_q^- - x_p^+} , \ &
    C^{ii}_{pq} &= (S^{ii}_{pq})^{-1}  \, \frac{x_q^+ - x_q^-}{x_q^- - x_p^+} \frac{\eta_p}{\eta_q} , \\
    F^{ii}_{pq} &= - (S^{ii}_{pq})^{-1} , \ &
    D^{ii}_{pq} &= (S^{ii}_{pq})^{-1}  \, \frac{x_q^- - x_p^-}{x_q^- - x_p^+} , \ &
    E^{ii}_{pq} &= (S^{ii}_{pq})^{-1}  \, \frac{x_p^+ - x_p^-}{x_q^- - x_p^+} \frac{\eta_q}{\eta_p} ,
  \end{aligned}
\end{equation}
where 
\begin{equation}
\eta_p=\sqrt{i(x_p^- - x_p^+)},
\end{equation}
and $S^{ii}_{pq}$ are undetermined scalar factors. By virtue of a discrete $\mathbb{Z}_2$ symmetry between left and right movers (``LR-symmetry''), we have that $S^{11}_{pq}=S^{\bar{1}\bar{1}}_{pq}$.

The form of the matrix elements~(\ref{eq:Sii-elements}) does not depend on the value of the mass $s$ in~(\ref{eq:zhukovski}). Therefore, when $i=3,\bar{3}$ they take the same form as before, up to replacing every $x^\pm\mapsto z^\pm$ and allowing for different phases $S^{33}_{pq}$ and $S^{\bar{3}\bar{3}}_{pq}$. Again, LR-simmetry imposes that $S^{33}_{pq}=S^{\bar{3}\bar{3}}_{pq}$. Furthermore, we expect that $S^{11}_{pq}$ and $S^{33}_{pq}$ have the same functional form up to specifying the value of the mass.


\subsection{The \texorpdfstring{$1\bar{1}$, $\bar{1}{1}$, $3\bar{3}$ and $\bar{3}{3}$}{11b, 1b1, 33b and 3b3} sectors}
When scattering particles having the same mass but opposite chirality, we find\footnote{Note that here we may have to insert or remove a vacuum site after the scattering. These length-changing effects are accounted for by a $Z^+$ or $Z^-$ insertion respectively.}
\begin{equation}\label{eq:T-ansatz-LR}
  \begin{aligned}
    \Smat \ket{\fixedspaceL{\psi_p^{i}{\psi}_q^{\bar{\imath}}}{\phi_p^i {\phi}_q^{\bar{\imath}}}} 
    &= \fixedspaceR{A^{i\bar{\imath}}_{pq}}{A^{i\bar{\imath}}_{pq}} \ket{\fixedspaceL{{\psi}_q^{\bar{\imath}}\psi_p^i}{{\phi}_q^{\bar{\imath}}\phi_p^i}} + \fixedspaceR{A^{i\bar{\imath}}_{pq}}{B^{i\bar{\imath}}_{pq}} \ket{{\psi}_q^{\bar{\imath}} \psi_p^i Z^-}, \qquad &
    \Smat \ket{\fixedspaceL{\psi_p^i{\psi}_q^{\bar{\imath}}}{\phi_p^i{\psi}_q^{\bar{\imath}}}} 
    &= \fixedspaceR{A^{i\bar{\imath}}_{pq}}{C^{i\bar{\imath}}_{pq}} \ket{\fixedspaceL{{\psi}_q^{\bar{\imath}}\psi_p^i}{{\psi}_q^{\bar{\imath}}\phi_p^i}}, \\
    \Smat \ket{\fixedspaceL{\psi_p^i{\psi}_q^{\bar{\imath}}}{\psi_p^i{\psi}_q^{\bar{\imath}}}} 
    &= \fixedspaceR{A^{i\bar{\imath}}_{pq}}{E^{i\bar{\imath}}_{pq}} \ket{\fixedspaceL{{\psi}_q^{\bar{\imath}}\psi_p^i}{{\psi}_q^{\bar{\imath}}\psi_p^i}} + \fixedspaceR{A^{i\bar{\imath}}_{pq}}{F^{i\bar{\imath}}_{pq}} \ket{{\phi}_q^{\bar{\imath}}\phi_p^i Z^+}, \qquad &
    \Smat \ket{\fixedspaceL{\psi_p^i{\psi}_q^{\bar{\imath}}}{\psi_p^i{\phi}_q^{\bar{\imath}}}} 
    &= \fixedspaceR{A^{i\bar{\imath}}_{pq}}{D^{i\bar{\imath}}_{pq}} \ket{\fixedspaceL{{\psi}_q^{\bar{\imath}}\psi_p^i}{{\phi}_q^{\bar{\imath}}\psi_p^i}}.
  \end{aligned}
\end{equation}
where it is understood that $\bar{\bar{\imath}}=i$ and $i=1,3,\bar{1},\bar{3}$. Then for $i=1,\bar{1}$ the coefficients read
\begin{equation}\label{eq:T-solution-LR}
  \begin{aligned}
    A^{i\bar{\imath}}_{pq} &= +(S^{i\bar{\imath}}_{pq})^{-1} \frac{1-\frac{1}{x_p^+ x_q^-}}{\Omega^+_{pq}\,\Omega^-_{pq}}, \ &
    B^{i\bar{\imath}}_{pq} &= -(S^{i\bar{\imath}}_{pq})^{-1} \frac{\eta_p \eta_q}{x_p^- x_q^-} \frac{1}{\Omega^+_{pq}\,\Omega^-_{pq}}, \ &
    C^{i\bar{\imath}}_{pq} &= (S^{i\bar{\imath}}_{pq})^{-1} \frac{\Omega^-_{pq}}{\Omega^+_{pq}}, \\
    E^{i\bar{\imath}}_{pq} &= -(S^{i\bar{\imath}}_{pq})^{-1} \frac{1-\frac{1}{x_p^- x_q^+}}{\Omega^+_{pq}\,\Omega^-_{pq}}, \ &
    F^{i\bar{\imath}}_{pq} &= -(S^{i\bar{\imath}}_{pq})^{-1} \frac{\eta_p \eta_q}{x_p^+ x_q^+} \frac{1}{\Omega^+_{pq}\,\Omega^-_{pq}}, \ &
    D^{i\bar{\imath}}_{pq} &= (S^{i\bar{\imath}}_{pq})^{-1} \frac{\Omega^+_{pq}}{\Omega^-_{pq}},
  \end{aligned}
\end{equation}
where
\begin{equation}\label{eq:tau-LR-RL}
  \Omega^\pm_{pq} = \sqrt{1-\frac{1}{x_p^\pm x_q^\pm}} .
\end{equation}
LR-symmetry and unitarity impose that $S^{\bar{1}1}_{pq}=S^{1\bar{1}}_{pq}$. The equations for $3\bar{3}$ and $\bar{3}3$ can be found by replacing $x^\pm\mapsto z^\pm$, and we expect the scalar factors $S^{\bar{3}3}_{pq}=S^{3\bar{3}}_{pq}$ to have the same functional form as $S^{\bar{1}1}_{pq}$ and $S^{1\bar{1}}_{pq}$, up to the different value of the mass.

\subsection{The \texorpdfstring{$13$, $\bar{1}\bar{3}$, $31$ and $\bar{3}\bar{1}$}{13, 1b3b, 31 and 3b1b} sectors}
We now consider scattering between two excitations having the same chirality but different mass. The S-matrix in this case is similar to the one found above, but for later convenience we will use a different normalization of the S-matrix elements, where
\begin{equation}
  \begin{aligned}
    \Smat \ket{\fixedspaceL{\psi^{i}_p\psi^j_q}{\phi^{i}_p\phi^j_q}} 
    &= \fixedspaceR{A^{ij}_{pq}}{A^{ij}_{pq}} \ket{\fixedspaceR{\psi^j_q\psi^{i}_p}{\phi^j_q\phi^{i}_p}} , \qquad &
    \Smat \ket{\fixedspaceL{\phi^{i}_p\psi^j_q}{\phi^{i}_p\psi^j_q}} 
    &= \fixedspaceR{A^{ij}_{pq}}{B^{ij}_{pq}} \ket{\fixedspaceR{\psi^j_q\phi^{i}_p}{\psi^j_q\phi^{i}_p}} 
    +  \fixedspaceR{A^{ij}_{pq}}{C^{ij}_{pq}} \ket{\fixedspaceR{\psi^j_q\phi^{i}_p}{\phi^j_q\psi^{i}_p}} , \\
    \Smat \ket{\fixedspaceL{\psi^{i}_p\psi^j_q}{\psi^{i}_p\psi^j_q}} 
    &= \fixedspaceR{A^{ij}_{pq}}{F^{ij}_{pq}} \ket{\fixedspaceR{\psi^j_q\psi^{i}_p}{\psi^j_q\psi^{i}_p}} , \qquad &
    \Smat \ket{\fixedspaceL{\phi^{i}_p\psi^j_q}{\psi^{i}_p\phi^j_q}} 
    &= \fixedspaceR{A^{ij}_{pq}}{D^{ij}_{pq}} \ket{\fixedspaceR{\psi^j_q\phi^{i}_p}{\phi^j_q\psi^{i}_p}}
    +  \fixedspaceR{A^{ij}_{pq}}{E^{ij}_{pq}} \ket{\fixedspaceR{\psi^j_q\phi^{i}_p}{\psi^j_q\phi^{i}_p}} ,
  \end{aligned}
\end{equation}
with $ij=13,31,\bar{1}\bar{3}$ or $\bar{3}\bar{1}$. Let us consider first the case where $ij=13$ or $\bar{1}\bar{3}$. Then
\begin{equation}
  \begin{aligned}
    A^{ij}_{pq} &= +(S^{ij}_{pq})^{-1}  , \quad &
    B^{ij}_{pq} &= (S^{ij}_{pq})^{-1}  \frac{z_q^+ - x_p^+}{z_q^+ - x_p^-} , \quad &
    C^{ij}_{pq} &= (S^{ij}_{pq})^{-1}  \frac{z_q^+ - z_q^-}{z_q^+ - x_p^-} \frac{\eta_{p}}{\eta_{q}} , \\
    F^{ij}_{pq} &= - (S^{ij}_{pq})^{-1}  \frac{z_q^- - x_p^+}{z_q^+ - x_p^-}, &
    D^{ij}_{pq} &= (S^{ij}_{pq})^{-1}  \frac{z_q^- - x_p^-}{z_q^+ - x_p^-} , &
    E^{ij}_{pq} &= (S^{ij}_{pq})^{-1}  \frac{x_p^+ - x_p^-}{z_q^+ - x_p^-} \frac{\eta_{q}}{\eta_{p}} ,
  \end{aligned}
\end{equation}
where once again the antisymmetric scalar factors satisfy $S^{13}_{pq}=S^{\bar{1}\bar{3}}_{pq}$. Again, to obtain the two remaining sectors it is enough to account for the values of the mass by swapping $x^\pm\leftrightarrow z^\pm$. In this case the scalar factors are related by unitarity.

\subsection{The \texorpdfstring{$1\bar{3}$, $\bar{1}3$, ${3}\bar{1}$ and $\bar{3}{1}$}{13b, 1b3, 31b and 3b1} sectors}
Finally, let us consider the scattering of excitations with different mass and chirality,
\begin{equation}
  \begin{aligned}
    \Smat \ket{\fixedspaceR{\psi_p^i{\psi}^{\bar{\jmath}}_q}{\phi_p^i{\phi}^{\bar{\jmath}}_q}} 
    &= \fixedspaceR{A^{i\bar{\jmath}}_{pq}}{A^{i\bar{\jmath}}_{pq}} \ket{\fixedspaceR{{\psi}^{\bar{\jmath}}_q\psi_p^i}{{\phi}^{\bar{\jmath}}_q\phi_p^i}}  
    +  \fixedspaceR{A^{i\bar{\jmath}}_{pq}}{B^{i\bar{\jmath}}_{pq}} \ket{\fixedspaceR{{\psi}^{\bar{\jmath}}_q\psi_p^i Z^-}{{\psi}^{\bar{\jmath}}_q\psi_p^i Z^-}} , \quad &
    \Smat \ket{\fixedspaceR{\phi_p^i{\psi}^{\bar{\jmath}}_q}{\phi_p^i{\psi}^{\bar{\jmath}}_q}} 
    &= \fixedspaceR{A^{i\bar{\jmath}}_{pq}}{C^{i\bar{\jmath}}_{pq}} \ket{\fixedspaceR{{\phi}^{\bar{\jmath}}_q\psi_p^i}{{\psi}^{\bar{\jmath}}_q\phi_p^i}} , \\
    \Smat \ket{\fixedspaceR{\psi_p^i{\psi}^{\bar{\jmath}}_q}{\psi_p^i{\psi}^{\bar{\jmath}}_q}} 
    &= \fixedspaceR{A^{i\bar{\jmath}}_{pq}}{E^{i\bar{\jmath}}_{pq}} \ket{\fixedspaceR{{\psi}^{\bar{\jmath}}_q\psi_p^i}{{\psi}^{\bar{\jmath}}_q\psi_p^i}} 
    +  \fixedspaceR{A^{i\bar{\jmath}}_{pq}}{F^{i\bar{\jmath}}_{pq}} \ket{\fixedspaceR{{\psi}^{\bar{\jmath}}_q\psi_p^i Z^-}{{\phi}^{\bar{\jmath}}_q\phi_p^i Z^+}} , \quad &
    \Smat \ket{\fixedspaceR{\phi_p^i{\psi}^{\bar{\jmath}}_q}{\psi_p^i{\phi}^{\bar{\jmath}}_q}} 
    &= \fixedspaceR{A^{i\bar{\jmath}}_{pq}}{D^{i\bar{\jmath}}_{pq}} \ket{\fixedspaceR{{\phi}^{\bar{\jmath}}_q\psi_p^i}{{\phi}^{\bar{\jmath}}_q\psi_p^i}}, 
  \end{aligned}
\end{equation}
with $ij=13,31,\bar{1}\bar{3}$ or $\bar{3}\bar{1}$ and $\bar{\bar{\jmath}}=j$. If we take $ij=13$ or $\bar{1}\bar{3}$ we have
\begin{equation}
  \begin{aligned}
    A^{i\bar{\jmath}}_{pq} &=  +(S^{i\bar{\jmath}}_{pq})^{-1} \frac{1-\frac{1}{x_p^+ z_q^-}}{\Omega^+_{pq}\,\Omega^-_{pq}} , \  &
    B^{i\bar{\jmath}}_{pq} &= -(S^{i\bar{\jmath}}_{pq})^{-1} \frac{\eta_{p} \eta_{q}}{x_p^- z_q^-}  \frac{1}{\Omega^+_{pq}\,\Omega^-_{pq}} , \  &
    C^{i\bar{\jmath}}_{pq} &=  (S^{i\bar{\jmath}}_{pq})^{-1} \frac{\Omega^-_{pq}}{\Omega^+_{pq}} , \\
    E^{i\bar{\jmath}}_{pq} &= -(S^{i\bar{\jmath}}_{pq})^{-1} \frac{1-\frac{1}{x_p^- z_q^+}}{\Omega^+_{pq}\,\Omega^-_{pq}}, \  &
    F^{i\bar{\jmath}}_{pq} &= -(S^{i\bar{\jmath}}_{pq})^{-1} \frac{\eta_{p} \eta_{q}}{x_p^+ z_q^+}  \frac{1}{\Omega^+_{pq}\,\Omega^-_{pq}} , \ &
    D^{i\bar{\jmath}}_{pq} &=  (S^{i\bar{\jmath}}_{pq})^{-1} \frac{\Omega^+_{pq}}{\Omega^-_{pq}},
  \end{aligned}
\end{equation}
where we, with a small abuse of notation, let $\Omega^\pm_{pq}$ depend on momenta through the appropriate Zhukovski parameterizations, \ie,
\begin{equation}
\Omega^\pm_{pq} = \sqrt{1-\frac{1}{x_p^\pm z_q^\pm}}.
\end{equation}
As before, the remaining blocks can be easily obtained by swapping $x^\pm\leftrightarrow z^\pm$, and the four scalar factors all take essentially the same form.  

\subsection{Crossing equations}
There are four distinct undetermined scalar factors in the S-matrix, playing a role akin to the one of the $\AdS_5\times \Sphere^5$ dressing phase. They are constrained by a set of crossing equations \cite{Borsato:2012ud}.\footnote{With respects to the first versions of this paper, we have rewritten the crossing equations in such a way that $p,q$ are on the real line and $\bar{q}$ is shifted upward by half of the imaginary period of the rapidity torus, following a standard convention.} For particles of the same mass these read, \eg,
\begin{equation}\label{eq:cross-11}
  S^{11}_{pq} \, S^{1\bar{1}}_{p\bar{q}} = \frac{x^-_p-x^+_q}{x^-_p-x^-_q}\sqrt{\frac{x^+_p}{x^-_p}\frac{x^-_p - x^-_q}{x^+_p - x^+_q}} , \qquad
  S^{11}_{p\bar{q}} \, S^{1\bar{1}}_{pq} = \frac{1-\frac{1}{x^+_px^+_q}}{1-\frac{1}{x^+_px^-_q}}\sqrt{\frac{1-\frac{1}{x^-_p x^-_q}}{1-\frac{1}{x^+_p x^+_q}}} ,
\end{equation}
and for different masses they are, \eg,
\begin{equation}
  \label{eq:cross-31}
  S^{31}_{pq} \, S^{3\bar{1}}_{p\bar{q}} = \frac{z^+_p-x^-_q}{z^-_p-x^-_q}\sqrt{\frac{z^+_p}{z^-_p}\frac{z^-_p - x^-_q}{z^+_p - x^+_q}} , \qquad
  S^{31}_{p\bar{q}} \, S^{3\bar{1}}_{pq} =\frac{z^+_p}{z^-_p}\frac{1-\frac{1}{z^+_px^+_q}}{1-\frac{1}{z^-_px^+_q}}\sqrt{\frac{1-\frac{1}{z^-_px^-_q}}{1-\frac{1}{z^+_px^+_q}}} .
\end{equation}
These functional equations involve one scalar factor evaluated at physical value of the momenta $p$ and $q$, and one where the rapidity corresponding to $q$ has been analytically continued to $\bar{q}$ so that $x_{\bar{q}}^\pm=1/x_q^\pm$ and the chirality of the corresponding excitation has been reversed.

The crossing equations are quite different from the one of $\AdS_5/\CFT_4$~\cite{Janik:2006dc}. However, taking the product of the two equations in~\eqref{eq:cross-11} we find a crossing equation for $S^{11}_{pq}  S^{1\bar{1}}_{p{q}}$ which takes a form resembling the one of $\AdS_5/\CFT_4$. Indeed the equation for the product can be solved at all-loop by
\begin{equation}\label{eq:crossing-product-sol}
  S^{11}_{pq} S^{1\bar{1}}_{pq}=\left(\frac{x^-_p}{x^+_p}\frac{x^+_q}{x^-_q}\right)^{1/4}\,\sigma_{\text{BES}}(p,q)\,,
\end{equation}
where $\sigma_{\text{BES}}$ is the BES phase~\cite{Beisert:2006ez}. This form of the phase is similar to what was recently found in $\AdS_3\times \Sphere^3\times\Torus^4$~\cite{Beccaria:2012kb}, as well as for Pohlmeyer reduced strings on $\AdS_3\times \Sphere^3$~\cite{Hoare:2011fj}.

\section{Diagonalisation of the S-matrix}
\label{sec:diagonalisation}
The S-matrix presented in the previous section satisfies the Yang-Baxter equation \cite{Borsato:2012ud} and is therefore compatible with factorized scattering. 
We can then use the two-particle S-matrix to construct eigenstates of the spin-chain Hamiltonian $\genH$. In this section we will construct \emph{asymptotic} eigenstates, living on a spin-chain of infinite length. In the next section we will adapt them to live on a finite periodic spin-chain, thereby obtaining a set of Bethe equations. 
An eigenstate of the spin-chain hamiltonian $\alg{H}$ containing $K$ excitations takes the form
\begin{equation}
\ket{\Psi}=\sum_{\pi\in S_K}\Smat_{\pi}\ket{\Psi}^{\I},
\end{equation}
where $\pi\in S_K$ is a permutation and $\ket{\Psi}^{\I}$ is a wavefunction. Thanks to the factorization of scattering, the $K$-body S-matrix can be written as a product of two-body S-matrices,
\begin{equation}
\Smat_{\pi} = \prod_{(k,l)\in\pi}\Smat_{kl},
\end{equation}
so that when acting on $\ket{\Psi}$ it permutes its excitations, 
\begin{equation}
\Smat_{\pi} \ket{\Psi} = S_{\pi}  \ket{\Psi}_{\pi}.
\end{equation}
From the form of the S-matrix we already know that some excitations do not scatter by a pure transmission of \emph{all} the quantum numbers among each other. We thus need the nesting procedure of~\cite{Beisert:2005wm,Beisert:2005tm,Beisert:2005fw,deLeeuw:2007uf} to perform the diagonalisation.

We call level-I vacuum $\ket{0}^{\I}$ the usual $\algSU(1|1)^2$ invariant one. Rather than considering all of its possible excitations at once, we will first restrict to a (maximal) set of excitations that scatter by transmitting \emph{all} quantum numbers and construct a  level-II vacuum $\ket{0}^{\II}$ out of these fields. The other fields are then interpreted as level-II excitations on top of this new vacuum. In principle one needs to repeat the procedure and restrict to those excitations that level by level transmit all the quantum numbers in a scattering process, until all the fields of the theory are accounted for. We will see that we only need to introduce two levels.

\paragraph{Level-I.}
The level-I vacuum is defined as $\ket{0}^{\I} \equiv \ket{Z^L}$ and all the fields can be viewed as excitations on this vacuum. This is the interpretation that was used in~\cite{Borsato:2012ud} to derive the S-matrix. Therefore, the S-matrix given in the previous section is the level-I S-matrix and we will denote it by $\Smat^{\I}$.

\paragraph{Level-II vacuum.}
Let us first choose a maximal set of excitations that transmit all the quantum numbers in a scattering process. From the form of $\Smat^{\I}$ we see that there are two possible choices,
\begin{equation}
  V^{\II}_A=\{ \phi^1, {\psi}^{\bar{1}}, \phi^3, {\psi}^{\bar{3}} \} \qquad \text{ or } \qquad  V^{\II}_B=\{ \psi^1 , {\phi}^{\bar{1}}, \psi^3, {\phi}^{\bar{3}} \}.
\end{equation}
Note that picking any particular excitation and letting it belong to the level-II vacuum (say, $\phi^1$) fixes this choice. For definiteness, we will use the set $V^{\II}_A$ to construct the level-II vacuum. This is consistent with the choice of $\genQ_{\smallL}$ and $\genS_{\smallR}$ as lowering operators. The Bethe equations that can be derived by choosing $V^{\II}_B$ are consistent with the choice of $\genQ_{\smallR}$ and $\genS_{\smallL}$ as lowering operators and are related to the previous Bethe equations by a fermionic duality (see section~\ref{sec:ferm-dual}).

We define the level-II vacuum as
\begin{equation}
  \ket{0}^{\II} = \ket{\mathcal{X}^{i_1}_{p_1} \mathcal{X}^{i_2}_{p_2} \cdots \mathcal{X}^{i_{K}}_{p_K}}^{\I},
\end{equation}
where $\mathcal{X}^{i_j}_{p_j}$ is one of the excitations in $V^{\II}_A$, and we use the index $i_j=1,3,\bar{1},\bar{3}$ to distinguish the different flavors. In total we have $K_1$ excitations of type~$\phi^1$, $K_{\bar{1}}$ excitations of type~${\psi}^{\bar{1}}$, $K_3$ of type~$\phi^3$ and~$K_{\bar{3}}$ of type~${\psi}^{\bar{3}}$, with $K_1 + K_{\bar{1}} + K_3 + K_{\bar{3}} = K$. By construction, for any permutation $\pi$ the S-matrix $\Smat^{\I}_{\pi}$ acts on the level-II vacuum as
\begin{equation}
  \Smat^{\I}_\pi \ket{0}^{\II} = S^{\I}_\pi \ket{0}^{\II}_\pi, \qquad
  \ket{0}^{\II}_\pi= \ket{\mathcal{X}^{i_{\pi(1)}}_{p_{\pi(1)}} \cdots \mathcal{X}^{i_{\pi(K)}}_{p_{\pi(K)}}}\!{}^{\I}.
\end{equation}
In the out-state the momenta, as well as the other quantum numbers, are permuted according to $\pi$. The phase $S^{\I}_\pi$ is given by the product of the S-matrix elements corresponding to the two-particle scatterings
\begin{equation}
S^{\I}_\pi = \prod_{(k,l)\in \pi} S_{i_k i_{l}}^{\I,\I}(p_k, p_{l}),
\end{equation}
where
\begin{equation}
  S_{ij}^{\I,\I}(p,q) = A^{ij}_{pq} , \qquad
  S_{i\bar{\jmath}}^{\I,\I}(p,q) = C^{i\bar{\jmath}}_{pq} , \qquad
  S_{\bar{\imath}j}^{\I,\I}(p,q) = D^{\bar{\imath}j}_{pq} , \qquad
  S_{\bar{\imath}\bar{\jmath}}^{\I,\I}(p,q) = F^{\bar{\imath}\bar{\jmath}}_{pq} ,
\end{equation}
for $i,j=1,3$ and $\bar{\imath},\bar{\jmath}=\bar{1},\bar{3}$.

\paragraph{Propagation.}

To take the other types of excitations into account, we can think of them as level-II excitations on the level-II vacuum. In order to do that we can act with the (super-)charges of the symmetry algebra on the states that compose the level-II vacuum. In particular, when we act with $\genQ_{\smallL}$ on $\ket{\phi^1}, \ket{{\psi}^{\bar{1}}}, \ket{\phi^3}, \ket{{\psi}^{\bar{3}}}$, we obtain respectively $\ket{\psi^1}, \ket{{\phi}^{\bar{1}} Z^+}, \ket{\psi^3}, \ket{{\phi}^{\bar{3}} Z^+}$. We will use $y$ to parameterize the corresponding level-II excitations.
Similarly, the action of $\genS_{\smallR}$ will create respectively $\ket{\psi^1 Z^-}, \ket{{\phi}^{\bar{1}}}, \ket{\psi^3 Z^-}, \ket{{\phi}^{\bar{3}}}$. The corresponding level-II excitations are now parametrized by $\bar{y}$.
In what follows we will illustrate the diagonalization of the S-matrix when considering the lowering operator $\genQ_{\smallL}$ only. The calculations for $\genS_{\smallR}$ are the same up to exchanging the left and right flavors. 

If we denote the excitation generated by the action of $\genQ_{\smallL}$ on $\ket{\mathcal{X}^{i_k}_{p_k}}$ by $\ket{\mathcal{Y}^{i_k}_{p_k}}$, a level-II state containing a single excitation takes the form\footnote{%
  Whether $\ket{\mathcal{Y}^{i_k}_{p_k}}$ is bosonic or fermionic depends on the statistics of the level-II vacuum state from which it is created.%
} %
\begin{equation}
  \ket{\mathcal{Y}_y}^{\II} = \sum_{k=1}^K \chi_k (y) \ket{\mathcal{X}^{i_1}_{p_1} \cdots \mathcal{Y}^{i_k}_{p_k} \cdots \mathcal{X}^{i_{K}}_{p_K}}^{\I}.
\end{equation}
We make a plane wave ansatz for the wave function $\chi_k (y)$ and write
\begin{equation}
  \chi_k (y) =f_{i_k}(y,p_k) \prod_{l=1}^{k-1} S_{i_k i_{l}}^{\II,\I} (y,p_{l}).
\end{equation}
The function $S_{i_k i_{l}}^{\II,\I}(y,p_{l})$ represents the scattering between the level-II excitation of type $i_k$ and the level-I excitation of type $i_{l}$ and it is the coefficient that will appear in the Bethe equations to describe such an interaction.
The factor $f_{i_k}(y,p_k)$ is associated with the creation of the level-II excitation on top of the level-I field. 
These two functions are determined by imposing compatibility of the S-matrix with the level-II excitation,
\begin{equation}
  \Smat^{\I}_\pi \ket{\mathcal{Y}_y}^{\II} = S^{\I}_\pi \ket{\mathcal{Y}_y}^{\II}_\pi, \qquad
  \ket{\mathcal{Y}_y}^{\II}_{\pi} = \sum_{k=1}^K \chi_{\pi,k} (y) \ket{\mathcal{X}^{i_{\pi{(1)}}}_{p_{\pi(1)}} \cdots \mathcal{Y}^{i_{\pi(k)}}_{p_{\pi(k)}} \cdots \mathcal{X}^{i_{\pi{(K)}}}_{p_{\pi(K)}}}^{\I},
\end{equation}
where the wave function for the permuted state is given by
\begin{equation}
\chi_{\pi,k} (y) = f_{i_{\pi(k)}}(y,p_{\pi(k)}) \prod_{l=1}^{k-1} S_{i_{\pi(k)} i_{\pi(l)}}^{\II,\I} (y,p_{\pi(l)}).
\end{equation}

To perform the calculations it is enough to consider a state containing only two excitations. We will consider separately left and right level-II excitations and their interaction with the left and right fields of the level-II vacuum. 
To make the notation simpler we will just present the calculations in the case of excitations of mass $\alpha$, as the generalization to mass $1-\alpha$ and different masses is straightforward.
Starting from a level-II vacuum defined as $\ket{0}^{\II}_{{11}}=\ket{\phi^{1}_{p} \phi^{1}_{q}}$ we can write
\begin{equation}
  \begin{aligned}
    \ket{\mathcal{Y}_y}^{\II}_{11} &= f_{1}(y,p) \ket{\psi^{1}_{p} \phi^{1}_{q}} + f_{1}(y,q) S_{{11}}^{\II,\I}(y,p) \ket{\phi^{1}_{p} \psi^{1}_{q}}, \\
    \ket{\mathcal{Y}_y}^{\II}_{11, \pi} &= f_{1}(y,q) \ket{\psi^{1}_{q} \phi^{1}_{p}} + f_{1}(y,p) S_{{11}}^{\II,\I}(y,q) \ket{\phi^{1}_{q} \psi^{1}_{p}},
  \end{aligned}
\end{equation}
where we use the subscript ``11'' to indicate the level-II vacuum.
The compatibility equation
\begin{equation}
  \Smat^{\I}_\pi \ket{\mathcal{Y}_y}^{\II}_{11} = A^{{11}}_{p q} \ket{\mathcal{Y}_y}^{\II}_{11, \pi}
\end{equation}
is solved by
\begin{equation}
  f_{1}(y,p) = g_{1}(y) \frac{\eta_p}{h_{1}(y) - x_p^+} , \qquad
  S_{{11}}^{\II,\I}(y,p) = \frac{h_{1}(y) - x_p^-}{h_{1}(y) - x_p^+},
\end{equation}
where $h_{1}(y), g_{1}(y)$ are arbitrary functions of $y$.

We can repeat the same procedure for right excitations that scatter with right fields of the level-II vacuum. Starting from $\ket{0}^{\II}_{{\bar{1}\bar{1}}}=\ket{{\psi}^{\bar{1}}_{p} {\psi}^{\bar{1}}_{q}}$, the two-particle states are
\begin{equation}
  \begin{aligned}
    \ket{\mathcal{Y}_y}^{\II}_{\bar{1}\bar{1}} &= f_{\bar{1}}(y,p) \ket{{\phi}^{\bar{1}}_{p} Z^+ {\psi}^{\bar{1}}_{q}} + f_{\bar{1}}(y,q) S_{{\bar{1}\bar{1}}}^{\II,\I}(y,p) \ket{{\psi}^{\bar{1}}_{p} {\phi}^{\bar{1}}_{q} Z^+}, \\
    \ket{\mathcal{Y}_y}^{\II}_{\bar{1}\bar{1}, \pi} &= f_{\bar{1}}(y,q) \ket{{\phi}^{\bar{1}}_{q} Z^+ {\psi}^{\bar{1}}_{p}} + f_{\bar{1}}(y,p) S_{{\bar{1}\bar{1}}}^{\II,\I}(y,q) \ket{{\psi}^{\bar{1}}_{q} {\phi}^{\bar{1}}_{p} Z^+}.
  \end{aligned}
\end{equation}
We now impose the equation
\begin{equation}
  \Smat^{\I}_\pi \ket{\mathcal{Y}_y}^{\II}_{\bar{1}\bar{1}} = F^{{\bar{1}\bar{1}}}_{p q} \ket{\mathcal{Y}_y}^{\II}_{\bar{1}\bar{1}, \pi},
\end{equation}
which is solved by
\begin{equation}
  f_{\bar{1}}(y,p) = \frac{-i g_{\bar{1}}(y)}{x_p^+} \frac{\eta_p}{1 - \frac{1}{h_{\bar{1}}(y) \ x_p^-}} , \qquad
  S_{{\bar{1}\bar{1}}}^{\II,\I}(y,p) =- \frac{1 - \frac{1}{h_{\bar{1}}(y) \ x_p^+}}{1 - \frac{1}{h_{\bar{1}}(y) \  x_p^-}}.
\end{equation}
As before $h_{\bar{1}}(y), g_{\bar{1}}(y)$ are generic functions of $y$.

If the level-II excitations are well defined, the above results should also be consistent with the case in which an excitation scatters with a level-II vacuum site of the opposite chirality. Let us consider for example the level-II vacuum $\ket{0}^{\II}_{{1 \bar{1}}} = \ket{\phi^{1}_{p} {\psi}^{\bar{1}}_{q}}$ and write
\begin{equation}
  \begin{aligned}
    \ket{\mathcal{Y}_y}^{\II}_{1\bar{1}} &= f_{1}(y,p) \ket{\psi^{1}_{p} \psi^{\bar{1}}_{q}} + f_{\bar{1}}(y,q) S_{{\bar{1} 1}}^{\II,\I}(y,p) \ket{\phi^{1}_{p} \phi^{\bar{1}}_{q} Z^+}, \\
    \ket{\mathcal{Y}_y}^{\II}_{1\bar{1}, \pi} &= f_{\bar{1}}(y,q) \ket{\phi^{\bar{1}}_{q} Z^+ \phi^{1}_{p}} + f_{1}(y,p) S_{{1 \bar{1}}}^{\II,\I}(y,q) \ket{\psi^{\bar{1}}_{q} \psi^{1}_{p}}.
  \end{aligned}
\end{equation}
The equation
\begin{equation}
\Smat^{\I}_\pi \ket{\mathcal{Y}_y}^{\II}_{1\bar{1}} = C^{{1 \bar{1}}}_{p q} \ket{\mathcal{Y}_y}^{\II}_{1\bar{1}, \pi}
\end{equation}
is solved by
\begin{equation}\label{eq:sol-nest-mix}
  \begin{aligned}
    h_{\bar{1}} (y) &=h_{1}(y) \equiv y, \qquad &
    g_{\bar{1}}(y) &=\frac{g_{1}(y)}{h_{1}(y)}, \\
    S_{{\bar{1} 1}}^{\II,\I}(y,p) &=S_{{11}}^{\II,\I}(y,p), \qquad &
    S_{{1 \bar{1}}}^{\II,\I}(y,p) &=S_{{\bar{1}\bar{1}}}^{\II,\I}(y,p).
  \end{aligned}
\end{equation}
Similar equations are valid when starting with the level-II vacuum $\ket{0}^{\II}_{{\bar{1} 1}}=\ket{{\psi}^{\bar{1}}_{p} \phi^{1}_{q}}$. The last line in (\ref{eq:sol-nest-mix}) can be considered a consistency check -- the level-II excitations produced by acting with $\genQ_{\smallL}$ have a level-II scattering matrix that only depends on the state of the level-II vacuum that they scatter with.

The above calculation does not make use of the mass of the involved excitations. Hence, the resulting S-matrix elements are the same if we change the type of one or more excitation from $1$ or $\bar{1}$ to $3$ or $\bar{3}$, provided we replace the parameters $x^\pm$ by $z^\pm$. We can also consider the excitations created by $\genS_{\smallR}$. Again the result is the same, up to the fact that we need to exchange left- and right-movers and replace the parameter $y$ by $\bar{y}$.

\paragraph{Scattering.}

We now consider a state with two level-II excitations
\begin{equation}
  \ket{\mathcal{Y}_{y_1} \mathcal{Y}_{y_2}}^{\II} = \sum_{\substack{k,l=1\\k<l}}^K \chi_k(y_1) \chi_l(y_2) 
  \ket{\mathcal{X}^{i_1}_{p_1} \dotsm \mathcal{Y}^{i_k}_{p_k} \dotsm \mathcal{Y}^{i_l}_{p_l} \dotsm \mathcal{X}^{i_{K}}_{p_K}}^{\I}.
\end{equation}
The above state satisfies the condition $\Smat_{\pi}^{\I} \ket{\mathcal{Y}_{y_1} \mathcal{Y}_{y_2}} = S_{\pi}^{\I} \ket{\mathcal{Y}_{y_1} \mathcal{Y}_{y_2}}_{\pi}$, except when the two level-II excitations sit on neighboring sites. It is therefore again enough to consider a state with $K=2$. Considering, \eg, the level-II vacuum $\ket{0}^{\II}_{{11}}=\ket{\phi^{1}_{p} \phi^{1}_{q}}$, we find the a two-excitation states
\begin{equation}
  \begin{aligned}
    \ket{\mathcal{Y}_{y_1} \mathcal{Y}_{y_2}}^{\II}_{11} &= f_1(y_1,p) f_1(y_2,q) S^{\II,\I}_{11}(y_2,p) \ket{\psi^1_p \psi^1_q} \\
    &\qquad + f_1(y_2,p) f_1(y_1,q) S^{\II,\I}_{11}(y_1,p) S^{\II,\II}_{11}(y_1,y_2) \ket{\psi^1_p \psi^1_q} , \\
    \ket{\mathcal{Y}_{y_1} \mathcal{Y}_{y_2}}^{\II}_{11,\pi} &= f_1(y_1,q) f_1(y_2,p) S^{\II,\I}_{11}(y_2,q) \ket{\psi^1_q \psi^1_p} \\
    &\qquad + f_1(y_2,q) f_1(y_1,p) S^{\II,\I}_{11}(y_1,q) S^{\II,\II}_{11}(y_1,y_2) \ket{\psi^1_q \psi^1_p} .
  \end{aligned}
\end{equation}
Demanding, as before, the compatibility condition
\begin{equation}
  \Smat^{\I}_\pi \ket{\mathcal{Y}_{y_1} \mathcal{Y}_{y_2}}^{\II}_{11} = A^{{11}}_{p q} \ket{\mathcal{Y}_{y_1} \mathcal{Y}_{y_2}}^{\II}_{11, \pi}
\end{equation}
to be satisfied, we find that the S-matrix element for the scattering of two level-II excitations is given by
\begin{equation}
  S^{\II,\II}(y_1,y_2) = -1,
\end{equation}
so that the excitations scatter trivially. It is straightforward to check that the above result is independent of what level-II vacuum we start with. We have therefore dropped the ``11'' subscript on $S^{\II,\II}$.

If we instead consider the case where one or both excitations are created by $\genS_{\smallR}$ we again find that the corresponding matrix elements $S^{\II,\II}(y,\bar{y})$ and $S^{\II,\II}(\bar{y}_1,\bar{y}_2)$ are trivial.

\section{Bethe Equations}
\label{sec:BA-equations}

In the following we denote the length of the spin-chain by $L$. We can express $L$ as
\begin{equation}
  \begin{aligned}
    L = N(Z) &+ \left( N(\fixedspaceR{\phi^1}{\phi^1}) + N(\fixedspaceR{\phi^1}{\phi^3}) + N(\fixedspaceR{\phi^1}{\psi^{\bar{1}}}) + N(\fixedspaceR{\phi^1}{\psi^{\bar{3}}}) \right) \\
    &+ \left( N(\fixedspaceR{\phi^1}{\psi^1}) + N(\fixedspaceR{\phi^1}{\psi^3}) + N(\fixedspaceR{\phi^1}{\phi^{\bar{1}}}) + N(\fixedspaceR{\phi^1}{\phi^{\bar{3}}}) \right) ,
  \end{aligned}
\end{equation}
where $N(Z)$ is the number of vacuum sites and $N(\mathcal{X})$ is the number of excitations of the corresponding type.

Thanks to the diagonalization procedure and the factorized scattering we can act with the S-matrix on an eigenstate $\ket{\Psi}$ and obtain a new state that is just proportional to the one in which all the excitations are permuted: the proportionality factor is given by the product of all the S-matrix elements of the pairwise scatterings.
Note that each scattering must be performed at all the levels, namely $\Smat \ket{\mathcal{X}_{p_1} \mathcal{Y}_{p_2}} = \prod_{B=0}^{\II} S^{A,B}(x_1^A, x_2^B) \ket{\mathcal{Y}_{p_2} \mathcal{X}_{p_1}}$, where $A$ indicates the level of excitation $\mathcal{X}$ and $B$ the level of $\mathcal{Y}$. In particular, when a level-I excitation scatters with a level-0 (vacuum) one, we get a phase $S^{\I,0}(x_j,\cdot) = x_{p_j}^+ /x_{p_j}^-=e^{i p_j}$.
The Bethe equations can thus be written as
\begin{equation}
1=\prod_{\substack{j = 1\\j \neq k}}^{L} \prod_{B_j =0}^{\II} S^{A,B_j}(x^A_k, x^{B_j}_{j}).
\end{equation}
The total momentum of physical states should vanish \ie, they should be annihilated by the central charges $\genP, \genK$~\cite{Borsato:2012ud}. This is the level-matching condition, which can be written as
\begin{equation}
\prod_j^{K_1} \frac{x_j^+}{x_j^-} \ \prod_j^{K_3} \frac{z_j^+}{z_j^-} \  \prod_j^{K_{\bar{1}}} \frac{\bar{x}_j^+}{\bar{x}_j^-} \ \prod_j^{K_{\bar{3}}} \frac{\bar{z}_j^+}{\bar{z}_j^-}=1.
\end{equation}
In the following, all the 16 scalar factors $S^{ij}$ will appear. Recall that they are related among themselves by LR-symmetry  and unitarity as explained in section~\ref{sec:S-matrix-recap}, but we will not make these constraints explicit here.

Once the level matching condition and Bethe ansatz equations are imposed, the energy of a multi-excitation state can be found from the momenta of its constituents  through the dispersion relation
\begin{equation}
\label{eq:dispersion}
E=E_{1}+E_{\bar{1}}+E_{3}+E_{\bar{3}},\qquad E_j=\sum_{k=1}^{K_j}\sqrt{s_j^2+4h^2\sin^2\frac{p_k}{2}} ,
\end{equation}
where $s_{1}=s_{\bar{1}}=\alpha$ and $s_{3}=s_{\bar{3}}=1-\alpha$.
The all-loop Bethe ansatz equations then take the form
\begin{align}\label{eq:bethe-equations}
    \begin{split}
      \left(\frac{x_k^+}{x_k^-}\right)^L &=
      \prod_{\substack{j = 1\\j \neq k}}^{K_1} \frac{x_k^+ - x_j^-}{x_k^- - x_j^+} S^{11}(x_k,x_j)
      \prod_{j=1}^{K_2} \frac{x_k^- - y_j}{x_k^+ - y_j}
      \prod_{j=1}^{K_3} S^{13}(x_k,z_j) 
      \\ &\phantom{\ = \ }\times
      \prod_{j=1}^{K_{\bar{1}}} \sqrt{\frac{1- \frac{1}{x_k^+ \bar{x}_j^+}}{1- \frac{1}{x_k^- \bar{x}_j^-}}} S^{1\bar{1}}(x_k,\bar{x}_j)
      \prod_{j=1}^{K_{\bar{2}}} \frac{1 - \frac{1}{x_k^- \bar{y}_j}}{1- \frac{1}{x_k^+ \bar{y}_j}}
      \prod_{j=1}^{K_{\bar{3}}} \sqrt{\frac{1- \frac{1}{x_k^+ \bar{z}_j^+}}{1- \frac{1}{x_k^- \bar{z}_j^-}}} S^{1\bar{3}}(x_k,\bar{z}_j) ,
    \end{split} \\
    1 &= 
    \prod_{j=1}^{K_1} \frac{y_k - x_j^+}{y_k - x_j^-}
    \prod_{j=1}^{K_3} \frac{y_k - z_j^+}{y_k - z_j^-}
    \prod_{j=1}^{K_{\bar{1}}} \frac{1 - \frac{1}{y_k \bar{x}_j^-}}{1- \frac{1}{y_k \bar{x}_j^+}}
    \prod_{j=1}^{K_{\bar{3}}} \frac{1 - \frac{1}{y_k \bar{z}_j^-}}{1- \frac{1}{y_k \bar{z}_j^+}} , \\
    \begin{split}
      \left(\frac{z_k^+}{z_k^-}\right)^L &=
      \prod_{\substack{j = 1\\j \neq k}}^{K_3} \frac{z_k^+ - z_j^-}{z_k^- - z_j^+}S^{33}(z_k, z_j)
      \prod_{j=1}^{K_2} \frac{z_k^- - y_j}{z_k^+ - y_j}
      \prod_{j=1}^{K_1} S^{31}(z_k,x_j) 
      \\ &\phantom{\ = \ }\times
      \prod_{j=1}^{K_{\bar{3}}} \sqrt{\frac{1- \frac{1}{z_k^+ \bar{z}_j^+}}{1- \frac{1}{z_k^- \bar{z}_j^-}}} S^{3\bar{3}}(z_k,\bar{z}_j)
      \prod_{j=1}^{K_{\bar{2}}} \frac{1 - \frac{1}{z_k^- \bar{y}_j}}{1- \frac{1}{z_k^+ \bar{y}_j}}
      \prod_{j=1}^{K_{\bar{1}}} \sqrt{\frac{1- \frac{1}{z_k^+ \bar{x}_j^+}}{1- \frac{1}{z_k^- \bar{x}_j^-}}} S^{3\bar{1}}(z_k,\bar{x}_j) ,
    \end{split}
\\
%
    \begin{split}
      \left(\frac{\bar{x}_k^+}{\bar{x}_k^-}\right)^L &=
      \prod_{\substack{j = 1\\j \neq k}}^{K_{\bar{1}}} S^{\bar{1}\bar{1}}(\bar{x}_k,\bar{x}_j)
      \prod_{j=1}^{K_{\bar{2}}} \frac{\bar{x}_k^+ - \bar{y}_j}{\bar{x}_k^- - \bar{y}_j}
      \prod_{j=1}^{K_{\bar{3}}} \frac{\bar{x}_k^- - \bar{z}_j^+}{\bar{x}_k^+ - \bar{z}_j^-} S^{\bar{1}\bar{3}}(\bar{x}_k,\bar{z}_j)
      \\ &\phantom{\ = \ }\times
      \prod_{j=1}^{K_1} \sqrt{\frac{1- \frac{1}{\bar{x}_k^- x_j^-}}{1- \frac{1}{\bar{x}_k^+ x_j^+}}} S^{\bar{1}1}(\bar{x}_k,x_j)
      \prod_{j=1}^{K_2} \frac{1 - \frac{1}{\bar{x}_k^+ y_j}}{1- \frac{1}{\bar{x}_k^- y_j}}
      \prod_{j=1}^{K_3} \sqrt{\frac{1- \frac{1}{\bar{x}_k^- z_j^-}}{1- \frac{1}{\bar{x}_k^+ z_j^+}}} S^{\bar{1}3}(\bar{x}_k,z_j) ,
    \end{split} \\
    1 &= 
    \prod_{j=1}^{K_{\bar{1}}} \frac{\bar{y}_k - \bar{x}_j^-}{\bar{y}_k - \bar{x}_j^+}
    \prod_{j=1}^{K_{\bar{3}}} \frac{\bar{y}_k - \bar{z}_j^-}{\bar{y}_k - \bar{z}_j^+}
    \prod_{j=1}^{K_1} \frac{1 - \frac{1}{\bar{y}_k x_j^+}}{1- \frac{1}{\bar{y}_k x_j^-}}
    \prod_{j=1}^{K_3} \frac{1 - \frac{1}{\bar{y}_k z_j^+}}{1- \frac{1}{\bar{y}_k z_j^-}} , \\
    \begin{split}
      \left(\frac{\bar{z}_k^+}{\bar{z}_k^-}\right)^L &=
      \prod_{\substack{j = 1\\j \neq k}}^{K_{\bar{3}}} S^{\bar{3}\bar{3}}(\bar{z}_k, \bar{z}_j)
      \prod_{j=1}^{K_{\bar{2}}} \frac{\bar{z}_k^+ - \bar{y}_j}{\bar{z}_k^- - \bar{y}_j}
      \prod_{j=1}^{K_{\bar{1}}} \frac{\bar{z}_k^- - \bar{x}_j^+}{\bar{z}_k^+ - \bar{x}_j^-} S^{\bar{3}\bar{1}}(\bar{z}_k,\bar{x}_j)
      \\ &\phantom{\ = \ }\times
      \prod_{j=1}^{K_3} \sqrt{\frac{1- \frac{1}{\bar{z}_k^- z_j^-}}{1- \frac{1}{\bar{z}_k^+ z_j^+}}} S^{\bar{3}3}(\bar{z}_k,z_j) 
      \prod_{j=1}^{K_2} \frac{1 - \frac{1}{\bar{z}_k^+ y_j}}{1- \frac{1}{\bar{z}_k^- y_j}}
      \prod_{j=1}^{K_1} \sqrt{\frac{1- \frac{1}{\bar{z}_k^- x_j^-}}{1- \frac{1}{\bar{z}_k^+ x_j^+}}} S^{\bar{3}1}(\bar{z}_k,x_j).
    \end{split}
\end{align}

\subsection{Fermionic duality}\label{sec:ferm-dual}
Under a fermionic duality~\cite{Essler:1992nk,Beisert:2005fw} the Bethe equations are modified but the spectrum of the theory remains invariant. 
The idea is that one can write a new set of equations equivalent to the previous one, in which all the auxiliary roots $y, \bar{y}$ are replaced by a dual set of auxiliary roots $\tilde{y}, \tilde{\bar{y}}$.
Here we highlight a nice feature of the Bethe equations, namely that a fermionic duality exchanges the left and right sectors.

Define the polynomial $P(\xi)$ as
\begin{equation}
\begin{aligned}
P(\xi) =& \prod_{j=1}^{K_1} (\xi - x_j^+)  \prod_{j=1}^{K_3} (\xi - z_j^+) \prod_{j=1}^{K_{\bar{1}}} (\xi - \frac{1}{ \bar{x}_j^-})  \prod_{j=1}^{K_{\bar{3}}} (\xi - \frac{1}{ \bar{z}_j^-}) \\
 & - \prod_{j=1}^{K_1} (\xi - x_j^-)  \prod_{j=1}^{K_3} (\xi - z_j^-)  \prod_{j=1}^{K_{\bar{1}}} (\xi- \frac{1}{ \bar{x}_j^+}) \prod_{j=1}^{K_{\bar{3}}} (\xi- \frac{1}{ \bar{z}_j^+}).
\end{aligned}
\end{equation}
This is a polynomial of degree $n=K_1 + K_3 + K_{\bar{1}} + K_{\bar{3}} -1$. Then the Bethe equations for excitations of type 2 and $\bar{2}$ can be rewritten respectively as
\begin{equation}
\begin{aligned}
P(y) &=0, & P(1/\bar{y})&=0.
\end{aligned}
\end{equation}
Another zero of the polynomial is at $\xi =0$ (the equation $P(0)=0$ is equivalent to the level-matching condition). We denote the remaining zeros by $\tilde{y}$ and $\tilde{\bar{y}}$, because they correspond to the dualization of respectevely 2 and $\bar{2}$ excitations. We can thus rewrite $P(\xi)$ as
\begin{equation}
P(\xi) = \xi \prod_{j=1}^{K_2} (\xi - y_j)  \prod_{j=1}^{\tilde{K}_2} (\xi - \tilde{y}_j) \prod_{j=1}^{K_{\bar{2}}} (\xi - \frac{1}{ \bar{y}_j})  \prod_{j=1}^{\tilde{K}_{\bar{2}}} (\xi - \frac{1}{ \tilde{\bar{y}}_j})
\end{equation}
for $\tilde{K}_2=K_1 + K_3 -K_2 -1$ and $\tilde{K}_{\bar{2}}=K_{\bar{1}} + K_{\bar{3}} -K_{\bar{2}} -1$ roots.
In the following we dualize the Bethe equations for excitations of type 1 and $\bar{1}$ (the procedure and the results are the same for 3 and $\bar{3}$).
We can now write the expression $P(x_k^+)/P(x_k^-)$ using the two possible representations for the polynomial, getting the equation
\begin{equation}
\begin{aligned}
 \frac{x_k^+}{x_k^-} \prod_{j=1}^{K_2} \frac{x_k^+ - y_j}{x_k^- - y_j}  \prod_{j=1}^{\tilde{K}_2} \frac{x_k^+ - \tilde{y}_j}{x_k^- - \tilde{y}_j} \prod_{j=1}^{K_{\bar{2}}} \frac{x_k^+ -1/ \bar{y}_j}{x_k^- - 1/ \bar{y}_j}  \prod_{j=1}^{\tilde{K}_{\bar{2}}} \frac{x_k^+ - 1/ \tilde{\bar{y}}_j}{x_k^- - 1/ \tilde{\bar{y}}_j} = \\
 - \prod_{j=1}^{K_1} \frac{x_k^+ - x_j^-}{x_k^- - x_j^+}  \prod_{j=1}^{K_3} \frac{x_k^+ - z_j^-}{x_k^- - z_j^+}  \prod_{j=1}^{K_{\bar{1}}} \frac{x_k^+ - 1/ \bar{x}_j^+}{x_k^- -1/ \bar{x}_j^-} \prod_{j=1}^{K_{\bar{3}}} \frac{x_k^+ - 1/ \bar{z}_j^+ }{x_k^- -1/\bar{z}_j^-}
\end{aligned}
\end{equation}
that becomes
\begin{equation}
\begin{aligned}
 \left(\frac{x_k^+}{x_k^-}\right)^{-1-K_{\bar{2}}-\tilde{K}_{\bar{2}}+K_{\bar{1}}+K_{\bar{3}}}  
\prod_{\substack{j = 1\\j \neq k}}^{K_1} \frac{x_k^+ - x_j^-}{x_k^- - x_j^+}  
\prod_{j=1}^{K_2} \frac{x_k^- - y_j}{x_k^+ - y_j} 
\prod_{j=1}^{K_{\bar{2}}} \frac{1 - \frac{1}{x_k^- \bar{y}_j}}{1 -\frac{1}{x_k^+ \bar{y}_j}} = \\
  \prod_{j=1}^{\tilde{K}_2} \frac{x_k^+ - \tilde{y}_j}{x_k^- - \tilde{y}_j}   
\prod_{j=1}^{\tilde{K}_{\bar{2}}} \frac{1 - \frac{1}{x_k^+ \tilde{\bar{y}}_j}}{1 - \frac{1}{x_k^- \tilde{\bar{y}}_j}} 
\prod_{j=1}^{K_3} \frac{x_k^- - z_j^+}{x_k^+ - z_j^-}  
\prod_{j=1}^{K_{\bar{1}}} \frac{1 -\frac{1}{x_k^- \bar{x}_j^-}}{1 - \frac{1}{x_k^+ \bar{x}_j^+}} 
\prod_{j=1}^{K_{\bar{3}}} \frac{1-\frac{1}{x_k^- \bar{z}_j^-}}{1 - \frac{1}{x_k^+ \bar{z}_j^+ }},
\end{aligned}
\end{equation}
where the exponent of $x_k^+/x_k^-$ is in fact 0. With the help of this substitution, the Bethe equation for type 1 excitations can thus be rewritten as
\begin{equation}
\begin{aligned}
      \left(\frac{x_k^+}{x_k^-}\right)^L = &
      \prod_{\substack{j = 1\\j \neq k}}^{K_1} S^{11}(x_k,x_j)
	 \prod_{j=1}^{\tilde{K}_2} \frac{x_k^+ - \tilde{y}_j}{x_k^- - \tilde{y}_j}
      \prod_{j=1}^{K_3} \frac{x_k^- - z_j^+}{x_k^+ - z_j^-} S^{13}(x_k,z_j)  \\ 
 &\times      \prod_{j=1}^{K_{\bar{1}}} \sqrt{\frac{1- \frac{1}{x_k^- \bar{x}_j^-}}{1- \frac{1}{x_k^+ \bar{x}_j^+}}} S^{1\bar{1}}(x_k,\bar{x}_j)
	\prod_{j=1}^{\tilde{K}_{\bar{2}}} \frac{1 - \frac{1}{x_k^+ \tilde{\bar{y}}_j}}{1 - \frac{1}{x_k^- \tilde{\bar{y}}_j}} 
      \prod_{j=1}^{K_{\bar{3}}} \sqrt{\frac{1- \frac{1}{x_k^- \bar{z}_j^-}}{1- \frac{1}{x_k^+ \bar{z}_j^+}}} S^{1\bar{3}}(x_k,\bar{z}_j).
\end{aligned}
\end{equation}
This equation has the same form of the original equation for $\bar{1}$ and is actually the same if one exchanges left and right.

Similarly, using $P(1/\bar{x}_k^+)/P(1/\bar{x}_k^-)$, one can obtain the dualized Bethe equation for type $\bar{1}$ excitations
\begin{equation}
\begin{aligned}
 \left(\frac{\bar{x}_k^+}{\bar{x}_k^-}\right)^L =&
      \prod_{\substack{j = 1\\j \neq k}}^{K_{\bar{1}}} \frac{\bar{x}_k^+ - \bar{x}_j^-}{\bar{x}_k^- - \bar{x}_j^+} S^{\bar{1}\bar{1}}(\bar{x}_k,\bar{x}_j)
      \prod_{j=1}^{\tilde{K}_{\bar{2}}} \frac{\bar{x}_k^- - \tilde{\bar{y}}_j}{\bar{x}_k^+ - \tilde{\bar{y}}_j}
      \prod_{j=1}^{K_{\bar{3}}} S^{\bar{1}\bar{3}}(\bar{x}_k,\bar{z}_j)      \\ 
&\times   \prod_{j=1}^{K_{1}} \sqrt{\frac{1- \frac{1}{\bar{x}_k^+ x_j^+}}{1- \frac{1}{\bar{x}_k^- x_j^-}}} S^{\bar{1}1}(\bar{x}_k,x_j)
      \prod_{j=1}^{\tilde{K}_{2}} \frac{1 - \frac{1}{\bar{x}_k^- \tilde{y}_j}}{1- \frac{1}{\bar{x}_k^+ \tilde{y}_j}}
      \prod_{j=1}^{K_{3}} \sqrt{\frac{1- \frac{1}{\bar{x}_k^+ z_j^+}}{1- \frac{1}{\bar{x}_k^- z_j^-}}} S^{\bar{1}3}(\bar{x}_k,z_j),
\end{aligned}
\end{equation}
which has the same form as the original Bethe equations for type 1 excitations.

After dualizing also the Bethe equations for 3 and $\bar{3}$ (with the same procedure and with similar results), one obtains a new set of Bethe equations. It is easy to see that such equations take the same form of the original ones up to exchanging left with right excitations and by substituting $(y,K_2)$ and $(\bar{y},K_{\bar{2}})$ by $(\tilde{y},\tilde{K}_2)$ and $(\tilde{\bar{y}},\tilde{K}_{\bar{2}})$, respectively.

\subsection{Small \texorpdfstring{$h$}{h} limit and Cartan matrix}
It is interesting to look at the weak-coupling expansion of the spin-chain. As explained in~\cite{Ogievetsky:1986hu,Minahan:2002ve,OhlssonSax:2011ms}, from the one-loop spin-chain BA we can read off the simple roots and the weights of the underlying $\alg{d}(2,1;\alpha)^2$ representation. In fact, we expect the BA equation for the $l$-th node  to take the form
\begin{equation}
  \label{eq:BE-one-loop}
  \left( \frac{u_{l,i} + \frac{i}{2} w_l}{u_{l,i} - \frac{i}{2} w_l} \right)^L
  = \prod_{\substack{k = 1\\k \neq i}}^{K_l} \frac{u_{l,i} - u_{l,k} + \frac{i}{2} A_{ll}}{u_{l,i} - u_{l,k} - \frac{i}{2} A_{ll}}
  \prod_{l' \neq l} \prod_{k=1}^{K_{l'}} 
  \frac{u_{l,i} - u_{l',k} + \frac{i}{2} A_{ll'}}{u_{l,i} - u_{l,k} - \frac{i}{2} A_{ll'}},
\end{equation}
where $w_l$ are weights and $A_{ll'}$ is an element of the Cartan matrix.
When $h\ll1$, let us expand
\begin{equation}
x^\pm\approx \frac{u_x\pm i\,\alpha}{h},\qquad
y\approx\frac{u_y}{h},\qquad
z^\pm\approx  \frac{u_z\pm i\,(1-\alpha)}{h},
\end{equation}
in the left sector, where $u_i$ are finite as $h\to0$, and similarly  in the right sector.

If we assume that the scalar factors $S^{ij}$ expand trivially in this limit, we indeed find that the Bethe ansatz takes the form~(\ref{eq:BE-one-loop}). Furthermore, the left and right sectors decouple, and the simple roots indeed correspond to the ones of $\alg{d}(2,1;\alpha)^2$. In fact it is immediate to read off the resulting Cartan matrix
\begin{equation}
\label{eq:cartan}
A=\left(
\begin{array}{cccccc}
 4 \alpha  & -2 \alpha  &0 &0 & 0 &0 \\
 -2 \alpha  & 0 & -2 (1-\alpha ) & 0 & 0 & 0 \\
0 & -2 (1-\alpha ) & 4 (1-\alpha ) &0 & 0 &0 \\
0 & 0 &0 & 0 & 2 \alpha  & -2 \\
 0 & 0 & 0 & 2 \alpha  & 0 & 2 (1-\alpha ) \\
0 & 0 &0 & -2 & 2 (1-\alpha ) & 0
\end{array}
\right),
\end{equation}
which is the one of $\alg{d}(2,1;\alpha)^2$ in the mixed grading of figure~\ref{fig:dynkin-d21a}. Furthermore, we find that the weights appearing in~\eqref{eq:BE-one-loop} are given by $w_1 = w_{\bar{1}} = 2\alpha$ and $w_3 = w_{\bar{3}} = 2(1-\alpha)$.
\begin{figure}
  \centering

  \subfloat[\label{fig:dynkin-d21a-orig}]{
    \begin{tikzpicture}
      \useasboundingbox (-1.1cm,-1.1cm) rectangle (1.6cm,1.1cm);

      \node (v1) at (-0.38cm,  0.65cm) [dynkin node] {};
      \node (v2) at ( 0.75cm,  0.00cm) [dynkin node] {};
      \node (v3) at (-0.38cm, -0.65cm) [dynkin node] {};

      \draw [dynkin line] (v2.south west) -- (v2.north east);
      \draw [dynkin line] (v2.north west) -- (v2.south east);

      \draw [dynkin line] (v1) -- (v2);
      \draw [dynkin line] (v2) -- (v3);

      \node at (v1.west) [anchor=east] {\small $1$};
      \node at (v3.west) [anchor=east] {\small $1$};
    \end{tikzpicture}
  }
  \hspace{1cm}
  \subfloat[\label{fig:dynkin-d21a-dual}]{
    \begin{tikzpicture}
      \useasboundingbox (-1.1cm,-1.1cm) rectangle (1.6cm,1.1cm);

      \node (v1) at (+0.38cm,  0.65cm) [dynkin node] {};
      \node (v2) at (-0.75cm,  0.00cm) [dynkin node] {};
      \node (v3) at (+0.38cm, -0.65cm) [dynkin node] {};

      \draw [dynkin line] (v1.south west) -- (v1.north east);
      \draw [dynkin line] (v1.north west) -- (v1.south east);

      \draw [dynkin line] (v2.south west) -- (v2.north east);
      \draw [dynkin line] (v2.north west) -- (v2.south east);

      \draw [dynkin line] (v3.south west) -- (v3.north east);
      \draw [dynkin line] (v3.north west) -- (v3.south east);

      \draw [dynkin line] (v1) -- (v2);
      \draw [dynkin line] (v2) -- (v3);
      \draw [dynkin line,double,double distance=0.5pt] (v3) -- (v1);

      \node at (v1.east) [anchor=west] {\small $1$};
      \node at (v3.east) [anchor=west] {\small $1$};
    \end{tikzpicture}
  }
  
  \caption{Two of the Dynkin diagrams for $\algD{\alpha}$. The crossed notes are fermionic and the labels indicate the momentum carrying roots in the Bethe equations. The $\algD{\alpha}^2$ Cartan matrix~\eqref{eq:cartan} corresponds to using diagram~\protect\subref{fig:dynkin-d21a-orig} for the left-movers and diagram~\protect\subref{fig:dynkin-d21a-dual} for the right-movers.}
  \label{fig:dynkin-d21a}
\end{figure}

The mixed grading did not appear in~\cite{Babichenko:2009dk,OhlssonSax:2011ms} and is here a result of the nesting procedure of section~\ref{sec:diagonalisation}. In fact, as explained in~\cite{Borsato:2012ud}, it can be seen as arising from our choice of the highest weight states of the left and right copies of $\alg{su}(1|1)$ used in the construction of the central extension~\cite{Borsato:2012ud}.

At the one-loop level, we have complete freedom in picking different simple roots for either copy of the algebra, and in particular to cast the Dynkin diagram in the form of~\cite{Babichenko:2009dk,OhlssonSax:2011ms}.\footnote{As we will see later, the same is true when constructing the finite gap equations, \ie, in the limit of large excitation numbers and strong coupling $h \gg 1$.}
However, there seems to be no way to do this in the all-loop Bethe ansatz. Indeed the fermionic duality of section~\ref{sec:ferm-dual} yields again equations corresponding to a mixed grading of the algebra, and the resulting Cartan matrix can be found by swapping the two diagonal blocks of~(\ref{eq:cartan}). It may still be possible that there exists some transformation on the all-loop equations that casts them in a different grading while preserving the spectrum, just as it happens at one-loop, but it is not clear which form it should take. This situation is similar to what happens in $\AdS_5/\CFT_4$, where such a transformation is also unknown in the all-loop case.

\subsection{Constraints on the scalar factors}
To fully determine the Bethe ansatz equations, we need to find the form of the scalar factors $S^{ij}$. As discussed above, this amounts to finding four functions (depending on whether the scattering occurs between excitations of the same or different masses and on the chiralities involved) constrained by the crossing equations~(\ref{eq:cross-11}) and~(\ref{eq:cross-31}).
In the case of $\AdS_5/\CFT_4$ and $\AdS_4/\CFT_3$ something similar happens, and the resulting dressing phases can be found in terms of the BES phase and simple functions of the Zhukovski variables~\cite{Beisert:2006ez,Beisert:2006ib,Gromov:2008qe,Volin:2009uv,Vieira:2010kb}. Here, this seems not to be the case.
Nonetheless, we can solve the crossing equations in the semiclassical limit. Knowing the pole structure arising in the near-BMN limit~\cite{Rughoonauth:2012qd} or in the finite gap equations~\cite{Babichenko:2009dk,Zarembo:2010yz}, we can solve~(\ref{eq:cross-11}) by the ansatz
\begin{align}
\label{eq:dressing-ansatz11}
S^{11}(x_1,x_2)&=\left(\frac{x^-_1}{x^+_1}\frac{x^+_2}{x^-_2}\right)^{1/2+\gamma_{1\bar{1}}}\left(\frac{1-\frac{1}{x^+_1 x^-_2}}{1-\frac{1}{x^-_1 x^+_2}}\right)^{1+2\gamma_{1\bar{1}}}\!\!\!\!\!\sigma^{2+4\gamma_{1\bar{1}}}(x_1,x_2),\\
S^{1\bar{1}}(x_1,\bar{x}_2)&=\left(\frac{x^-_1}{x^+_1}\frac{\bar{x}^+_2}{\bar{x}^-_2}\right)^{1/2+\gamma_{1\bar{1}}}\left(\frac{1-\frac{1}{x^+_1 \bar{x}^-_2}}{1-\frac{1}{x^-_1 \bar{x}^+_2}}\right)^{1/2+2\gamma_{1\bar{1}}}\!\!\!\!\!\!\!\!\!\!\sigma^{2+4\gamma_{1\bar{1}}}(x_1,\bar{x}_2) ,
\end{align}
where the function $\sigma(x_1,x_2)$ is an antisymmetric phase that reduces to the AFS one~\cite{Arutyunov:2004vx} in the semiclassical limit.\footnote{Requiring that this is compatible with~\eqref{eq:crossing-product-sol} would set $\gamma_{1\bar{1}}=-3/8$.}
Similar results hold for $S^{33},S^{3\bar{3}}$ after replacing $x^\pm\mapsto z^\pm$. We call $\gamma_{3\bar{3}}$ the corresponding undetermined coefficient.

We will assume that the equations for $S^{31}$ and $S^{3\bar{1}}$ can be solved in the semiclassical limit in terms of a suitable generalization of the AFS phase~\cite{Arutyunov:2004vx}, coupling particles of different masses. We therefore define
\begin{equation}
\label{eq:generalized-AFS}
\sigma(x_{1},x_{2})=\left(\frac{1-\frac{1}{x_{1}^- x_{2}^+}}{1-\frac{1}{x_{1}^+ x_{2}^-}}\right)
\left(\frac{1-\frac{1}{x_{1}^+ x_{2}^-}}{1-\frac{1}{x_{1}^+ x_{2}^+}}\frac{1-\frac{1}{x_{1}^- x_{2}^+}}{1-\frac{1}{x_{1}^- x_{2}^-}}\right)^{i\frac{h}{W_{12}}(x_{1}+1/x_{1}-x_{2}-1/x_{2})},
\end{equation}
where 
\begin{equation}
W_{12}=4\frac{s_1\,s_2}{s_1+s_2}=\begin{cases}
2\,s_1&\text{for}\quad s_1=s_2.\\
4\,s_1\, s_2&\text{for}\quad s_1+s_2=1.
\end{cases}
\end{equation}
For the case of excitations of different masses we solve equation~\eqref{eq:cross-31} by writing
\begin{align}
\label{eq:dressing-ansatz13}
  S^{31}(z_1,x_2) &= \left(
    \frac{z^-_1}{z^+_1}\frac{x^+_2}{x^-_2}
  \right)^{\gamma_{3\bar{1}}}
  \!\!
  \left(
    \frac{1-\frac{1}{z^+_1 x^-_2}}{1-\frac{1}{z^-_1 x^+_2}}
  \right)^{1+2\gamma_{3\bar{1}}}
  \sigma^{2+4\gamma_{3\bar{1}}}(z_1,x_2),\\
  S^{3\bar{1}}(z_1,\bar{x}_2) &= \left(
    \frac{z^-_1}{z^+_1}\frac{\bar{x}^+_2}{\bar{x}^-_2}
  \right)^{\gamma_{3\bar{1}}}
  \!\!
  \left(
    \frac{1-\frac{1}{z^+_1 \bar{x}^-_2}}{1-\frac{1}{z^-_1 \bar{x}^+_2}}
  \right)^{3/2+2\gamma_{3\bar{1}}}
  \sigma^{2+4\gamma_{3\bar{1}}}(z_1,\bar{x}_2).
\end{align}
The remaining scalar factors can be written down in a similar way, introducing several real constants $\gamma_{ij}$. As we have argued earlier, we expect the whole description of the spin-chain to be invariant under left-right symmetry which, as we have seen above, amounts to a fermionic duality on the BA equations. Furthermore, we assume that the real parameters in our ans\"atze do not depend explicitly on the mass, so that, \eg, $\gamma_{1\bar{1}}=\gamma_{3\bar{3}}$. Supplementing these requirements by unitarity, we can conclude that only two free parameters appear in the semiclassical limit of our Bethe ansatz. One coefficient $\gamma\equiv\gamma_{i\bar{\imath}}$ is common to all phases relating particles of the same mass, and the other $\Gamma\equiv\gamma_{i\bar{\jmath}}$ is common to the ones relating different masses.

This residual freedom is an artifact of our perturbative approach. Once suitable analytic properties for the scalar factors are assumed, we expect the crossing equations to have a unique set of solutions. However, this perturbative analysis already gives some interesting insight on the scalar factors. In particular, it is easy to check that the solution $\Gamma=-1/2$, where only simple phases couple the nodes of different masses,  is not a solution of the crossing equations to all-loop orders. This implies that a non-trivial coupling among such nodes has to be in the Bethe ansatz.

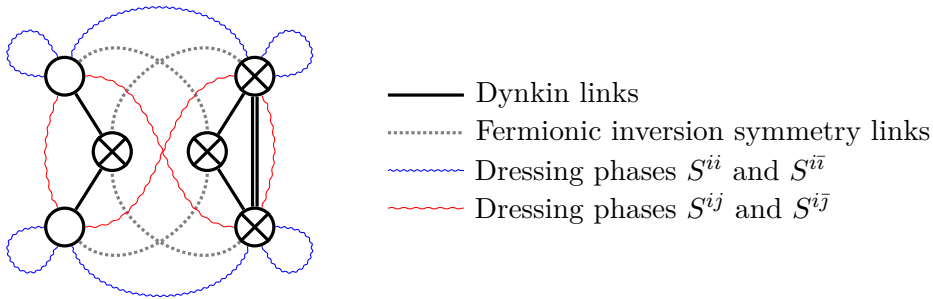
\begin{figure}
  \centering
  
\begin{tikzpicture}
  \begin{scope}
    \coordinate (m) at (0cm,0cm);

    \node (v1L) at (-1.25cm,   1cm) [dynkin node] {};
    \node (v2L) at (-0.625cm,  0cm) [dynkin node] {};
    \node (v3L) at (-1.25cm,  -1cm) [dynkin node] {};

    \draw [dynkin line] (v2L.south west) -- (v2L.north east);
    \draw [dynkin line] (v2L.north west) -- (v2L.south east);

    \draw [dynkin line] (v1L) -- (v2L);
    \draw [dynkin line] (v2L) -- (v3L);
    
    \node (v1R) at (+1.25cm,   1cm) [dynkin node] {};
    \node (v2R) at (+0.625cm,  0cm) [dynkin node] {};
    \node (v3R) at (+1.25cm,  -1cm) [dynkin node] {};

    \draw [inverse line] [out=  0+45,in= 90] (v1L) to (v2R);
    \draw [inverse line] [out=  0-45,in=270] (v3L) to (v2R);
    \draw [inverse line] [out=180-45,in= 90] (v1R) to (v2L);
    \draw [inverse line] [out=180+45,in=270] (v3R) to (v2L);

    \draw [dynkin line] (v1R.south west) -- (v1R.north east);
    \draw [dynkin line] (v1R.north west) -- (v1R.south east);

    \draw [dynkin line] (v2R.south west) -- (v2R.north east);
    \draw [dynkin line] (v2R.north west) -- (v2R.south east);

    \draw [dynkin line] (v3R.south west) -- (v3R.north east);
    \draw [dynkin line] (v3R.north west) -- (v3R.south east);

    \draw [dynkin line] (v1R) -- (v2R);
    \draw [dynkin line] (v2R) -- (v3R);
    \draw [dynkin line,double,double distance=0.5pt] (v3R) -- (v1R);

    \draw [red phase] [out=270-20,in= 90+20] (v1L) to (v3L);
    \draw [blue phase] [out=270+20,in=270-20] (v3L) to (v3R);
    \draw [red phase] [out= 90-20,in=270+20] (v3R) to (v1R);
    \draw [blue phase] [out= 90+20,in= 90-20] (v1R) to (v1L);

    \draw [red phase] [out=0,in=180] (v1L) to (v3R);
    \draw [red phase] [out=0,in=180] (v3L) to (v1R);

    \draw [blue phase] [out=180,in= 90+20,loop] (v1L) to (v1L);
    \draw [blue phase] [out=180,in=270-20,loop] (v3L) to (v3L);
    \draw [blue phase] [out=  0,in= 90-20,loop] (v1R) to (v1R);
    \draw [blue phase] [out=  0,in=270+20,loop] (v3R) to (v3R);
  \end{scope}

  \begin{scope}[xshift=+3cm,yshift=-0.75cm]
    \draw [dynkin line]  (0cm,1.5cm) -- (1cm,1.5cm) node [anchor=west,black] {\small Dynkin links};
    \draw [inverse line] (0cm,1.0cm) -- (1cm,1.0cm) node [anchor=west,black] {\small Fermionic inversion symmetry links};
    \draw [blue phase]   (0cm,0.5cm) -- (1cm,0.5cm) node [anchor=west,black] {\small Dressing phases $S^{ii}$ and $S^{i\bar{\imath}}$};
    \draw [red phase]    (0cm,0.0cm) -- (1cm,0.0cm) node [anchor=west,black] {\small Dressing phases $S^{ij}$ and $S^{i\bar{\jmath}}$};
  \end{scope}
\end{tikzpicture}

  \caption{The Dynkin diagram for $\algD{\alpha}^2$ in the mixed grading~\eqref{eq:cartan}, with the various interaction terms in~\eqref{eq:bethe-equations} indicated.}
  \label{fig:bethe-equations}
\end{figure}

The various couplings appearing in the Bethe equations are summarized in figure~\ref{fig:bethe-equations}.

\section{Semiclassical spectrum and comparisons}
\label{sec:comparison}
Here we will compare our all-loop BA with the finite gap~\cite{Babichenko:2009dk,Zarembo:2010yz} and near-BMN spectra~\cite{Rughoonauth:2012qd}. As a preliminary step, we will briefly present the construction of the finite gap equations for the supersymmetric coset with grading \eqref{eq:cartan}.

\subsection{Finite gap equations from supersymmetric coset}

In this section we repeat the finite gap construction of~\cite{Babichenko:2009dk,Zarembo:2010yz} to obtain a set of equations written in the mixed grading and to highlight some aspects of this construction. Note that the resulting equations are equivalent to those of~\cite{Babichenko:2009dk}, but written using a different grading of the algebra. The two gradings are related by a fermionic duality.\footnote{%
  The procedure is similar to the one discussed in section~\ref{sec:ferm-dual}, but at the level of finite gap equations it can be performed independently in the left- and right-moving sectors.%
}%

The finite gap (FG) equations describe a class of solutions of a classical integrable model. In our case, classical integrability immediately follows from the presence of an additional $\mathbb{Z}_4$ symmetry in the coset model. Each FG solution is described by a Riemann surface of finite genus, parameterized by a spectral parameter $x$. The eigenvalues of the monodromy matrix are related to a set of quasi-momenta $p_l(x)$ that take value on the Riemann sheets.
The quasi-momenta have poles at $x=\pm 1$ parameterized by

\begin{equation}
p_l(x) = \frac{1}{2} \frac{\kappa_l \pm 2 \pi m_l}{x \mp 1} + \cdots , \qquad  (x \rightarrow \pm 1).
\end{equation}
They also present discontinuities at the cuts of the Riemann sheets. The monodromies around the branching points $x_i$ are given by

\begin{equation}
p_l(x)  \rightarrow  p_l(x) + A_{lk} \, p_k(x) + 2 \pi n_{l,i},  \\
\end{equation}
where $A_{lk}$ are the elements of the Cartan matrix, which in our case is given by~\eqref{eq:cartan}.

The $\mathbb{Z}_4$ symmetry of the coset is implemented by the block-antidiagonal matrix $S$ that can be taken to be $S=\sigma_1\otimes \tilde{S}$, where in our grading
\begin{equation}
\tilde{S}=\pm\left(\begin{array}{ccc}
1&0&0\\
1&-1&1\\
0&0&1
\end{array}\right).
\end{equation}
The choice of the overall sign corresponds to a choice of the relative signs of the Cartan elements in the right moving sector with respect to the ones in the left moving one. In particular, the two possibilities are compatible with two ways of identifying the hamiltonian $\genH$ in terms of the left and right $\algD{\alpha}$ generators, namely
\begin{equation}
\label{eq:hamiltonian-def}
\genH = \genH_L \mp \genH_R .
\end{equation}
In order to compare the finite gap equations with our Bethe ansatz, the preferred choice is to pick a negative overall sign in the definition of $\tilde{S}$, consistent with the positive choice in the definiton of the spin-chain energy \eqref{eq:hamiltonian-def}.\footnote{The other choice would lead to an equivalent set of finite gap equations, which can also be obtained by flipping the sign of the denisities and mode numbers in the right-moving sector.}
The action of the $\mathbb{Z}_4$ symmetry is implemented on the quasi-momenta as $p_l(1/x)=S_{lm} \, p_m(x)$ that gives
\begin{equation}
S_{lk} \kappa_k =-\kappa_l, \qquad S_{lk} m_k=-m_l.
\end{equation}
We can then write the quasi-momenta in terms of densities $\rho_l(x)$ with support on the cuts as
\begin{equation}
p_l(x) = -\frac{\kappa_l x + 2 \pi m_l}{x^2 - 1} + \int \frac{\rho_l(y)}{x-y}dy - S_{lm} \int \frac{\rho_m(y)}{x-1/y} \frac{dy}{y^2}.
\end{equation}

The vector $\kappa$ can be found by imposing the Virasoro constraints, which translate into the null condition
\begin{equation}\label{eq:winding}
\begin{gathered}
0 = (\kappa_l\pm2\pi m_l)A_{lk}(\kappa_k \pm 2\pi m_k),\\
2\pi m_l = \left(\delta_{lk}-S_{lk}\right)\,\int \frac{dx}{x}\rho_k(x)\equiv\left(\delta_{lk}-S_{lk}\right)\,\mathcal{P}_k,
\end{gathered}
\end{equation}
where we define
\begin{equation}
\mathcal{P}_i=\int \frac{\rho_i(x)}{x}\,dx.
\end{equation}
The null condition is fulfilled if $\kappa=-2\pi \mathcal{E}(0,1,0,0,1,0)^t$ and if two distinct conditions on the momenta are satisfied
\begin{equation}
\label{eq:stronger-level-matching}
\mathcal{P}_1+\mathcal{P}_{\bar{1}}=0,\quad\quad\mathcal{P}_3+\mathcal{P}_{\bar{3}}=0.
\end{equation}
These two conditions are stronger that the usual level-mathcing condition, which reads 
\begin{equation}
\label{eq:weaker-level-matching}
0= \mathcal{P}_{\text{tot}}\equiv \alpha\,\left(\mathcal{P}_1+\mathcal{P}_{\bar{1}}\right)+(1-\alpha)\left(\mathcal{P}_3+\mathcal{P}_{\bar{3}}\right).
\end{equation}
The stronger condition~\eqref{eq:stronger-level-matching} is not due to our choice of the grading. A similar constraint appears to emerge from solving the null condition also in the original construction of~\cite{Babichenko:2009dk,Zarembo:2010yz}. 
Finally, the finite gap equations can be written as
\begin{equation}
2\pi n_l=-A_{lk}\frac{\kappa_k\, x+2\pi m_k}{x^2-1}+A_{lm}\pint \frac{\rho_m(y)}{x-y}dy-A_{lk}S_{km}\int\frac{\rho_m(y)}{x-1/y}\frac{dy}{y^2},
\end{equation}
where the spectral parameter takes values on the physical domain $|x|>1$ and the winding numbers $m_l$ are given by \eqref{eq:winding}.

Let us give the explicit form of the FG equations in our preferred grading~\eqref{eq:cartan},
\begin{equation}
\begin{aligned}
2\pi n_1&=-2\alpha\frac{x}{x^2-1}2\pi\mathcal{E}+4\alpha\pint\frac{\rho_1(y)}{x-y}dy-2\alpha\int\frac{\rho_2(y)}{x-y}dy \\
&\phantom{={}}+2\alpha\int\frac{\rho_{\bar{1}}(y)}{x-1/y}\frac{dy}{y^2}+2\alpha\int\frac{\rho_{\bar{2}}(y)}{x-1/y}\frac{dy}{y^2}-2\alpha\int\frac{\rho_{\bar{3}}(y)}{x-1/y}\frac{dy}{y^2} \\
&\phantom{={}}+2\alpha\frac{1}{x^2-1}(-2\mathcal{P}_1+\mathcal{P}_2-\mathcal{P}_{\bar{1}}-\mathcal{P}_{\bar{2}}+\mathcal{P}_{\bar{3}}), \\
2\pi n_2&=-2\alpha\int\frac{\rho_1(y)}{x-y}dy-2(1-\alpha)\int\frac{\rho_3(y)}{x-y}dy-2\alpha\int\frac{\rho_{\bar{1}}(y)}{x-1/y}\frac{dy}{y^2} \\
&\phantom{={}}-2(1-\alpha)\int\frac{\rho_{\bar{3}}(y)}{x-1/y}\frac{dy}{y^2}+2\frac{1}{x^2-1}(\alpha\mathcal{P}_1+(1-\alpha)\mathcal{P}_3+\alpha\mathcal{P}_{\bar{1}}+(1-\alpha)\mathcal{P}_{\bar{3}}), \\
2\pi n_3&=-2(1-\alpha)\frac{x}{x^2-1}2\pi\mathcal{E}-2(1-\alpha)\int\frac{\rho_2(y)}{x-y}dy+4(1-\alpha)\pint\frac{\rho_3(y)}{x-y}dy \\
&\phantom{={}}-2(1-\alpha)\int\frac{\rho_{\bar{1}}(y)}{x-1/y}\frac{dy}{y^2}+2(1-\alpha)\int\frac{\rho_{\bar{2}}(y)}{x-1/y}\frac{dy}{y^2}+2(1-\alpha)\int\frac{\rho_{\bar{3}}(y)}{x-1/y}\frac{dy}{y^2} \\
&\phantom{={}}+2(1-\alpha)\frac{1}{x^2-1}(\mathcal{P}_2-2\mathcal{P}_3+\mathcal{P}_{\bar{1}}-\mathcal{P}_{\bar{2}}-\mathcal{P}_{\bar{3}}), \\
2\pi n_{\bar{1}}&=-2\alpha\frac{x}{x^2-1}2\pi\mathcal{E}+2\alpha\int\frac{\rho_1(y)}{x-1/y}\frac{dy}{y^2}-2\alpha\int\frac{\rho_2(y)}{x-1/y}\frac{dy}{y^2} \\
&\phantom{={}}-2(1-\alpha)\int\frac{\rho_3(y)}{x-1/y}\frac{dy}{y^2}+2\alpha\int\frac{\rho_{\bar{2}}(y)}{x-y}dy-2\int\frac{\rho_{\bar{3}}(y)}{x-y}dy \\
&\phantom{={}}+2\frac{1}{x^2-1}(-\alpha\mathcal{P}_1+\alpha\mathcal{P}_2+(1-\alpha)\mathcal{P}_3-\alpha\mathcal{P}_{\bar{2}}+\mathcal{P}_{\bar{3}}), \\
2\pi n_{\bar{2}}&=+2\alpha\int\frac{\rho_1(y)}{x-1/y}\frac{dy}{y^2}+2(1-\alpha)\int\frac{\rho_3(y)}{x-1/y}\frac{dy}{y^2}+2\alpha\int\frac{\rho_{\bar{1}}(y)}{x-y}dy \\
&\phantom{={}}+2(1-\alpha)\int\frac{\rho_{\bar{3}}(y)}{x-y}dy-2\frac{1}{x^2-1}(\alpha\mathcal{P}_1+(1-\alpha)\mathcal{P}_3+\alpha\mathcal{P}_{\bar{1}}+(1-\alpha)\mathcal{P}_{\bar{3}}), \\
2\pi n_{\bar{3}}&=-2(1-\alpha)\frac{x}{x^2-1}2\pi\mathcal{E}-2\alpha\int\frac{\rho_1(y)}{x-1/y}\frac{dy}{y^2}-2(1-\alpha)\int\frac{\rho_2(y)}{x-1/y}\frac{dy}{y^2} \\
&\phantom{={}}+2(1-\alpha)\int\frac{\rho_3(y)}{x-1/y}\frac{dy}{y^2}-2\int\frac{\rho_{\bar{1}}(y)}{x-y}dy+2(1-\alpha)\int\frac{\rho_{\bar{2}}(y)}{x-y}dy \\
&\phantom{={}}+2\frac{1}{x^2-1}(\alpha\mathcal{P}_1+(1-\alpha)\mathcal{P}_2-(1-\alpha)\mathcal{P}_3+\mathcal{P}_{\bar{1}}-(1-\alpha)\mathcal{P}_{\bar{2}}). 
\end{aligned}
\end{equation}
Note that the winding contribution in the equations for $2$ and $\bar{2}$ is proportional to the total momentum $\mathcal{P}_{\text{tot}}$, and therefore vanishes upon imposing the level-matching condition~\eqref{eq:weaker-level-matching}. Similarly, by~\eqref{eq:weaker-level-matching} we can simplify the winding contribution  for $1,\bar{1}$ and $3,\bar{3}$ so that excitations of the same mass have the same winding. If we furthermore impose the stronger level matching condition~\eqref{eq:stronger-level-matching}, we also find that the winding for $1,\bar{1}$ and $3,\bar{3}$ equals to $ 2\alpha \mathcal{M}$ and  $ 2(1-\alpha) \mathcal{M}$ respectively, where
\begin{equation}
\mathcal{M}=-\frac{\mathcal{P}_{1}-\mathcal{P}_{\bar{1}}}{2}+\mathcal{P}_2-\mathcal{P}_{\bar{2}}-\frac{\mathcal{P}_{{3}}-\mathcal{P}_{\bar{3}}}{2}\,.
\end{equation}
It would be worth investigating which one of the two conditions~\eqref{eq:stronger-level-matching} and~\eqref{eq:weaker-level-matching} holds in string theory, for instance by explicitly constructing classical solutions with non-trivial winding. 

\subsection{Comparison with finite gap limit}
The finite gap equations that we found by classical integrability can also be thought of as the semiclassical limit of the all-loop Bethe equations. Let us take the spin-chain to be long and consider the case where both the coupling constant $h$ and the number of excitations $K_i$ are large. Then, since $h\gg1$, the quantum fluctuations of the string are suppressed. In the thermodynamic limit $L\approx K_i\gg1$ we expect the Bethe roots to condense on the cuts that appear in the finite gap equations. Strictly speaking, one would need to prove that this is the case, as it was done \eg in~\cite{Beisert:2005di} for $\AdS_5/\CFT_4$. Here we shall assume so for the purpose of comparing the form of the resulting equations.

Let us then take the semiclassical limit of our all-loop conjectured Bethe ansatz. Following a standard route, we take the densities to be given, in terms of the Bethe roots, by
\begin{equation}
\rho_i(x)=\sum_{k=1}^{K_i}\frac{x^2}{x^2-1}\delta(x-x_{i,k}),\quad\quad i=1,2,3,\bar{1},\bar{2},\bar{3}.
\end{equation}
where the excitation numbers are large $K_i\gg1$ and we make use of the expansion
\begin{equation}
x_i^\pm \approx x_i\pm i\frac{s_i}{h}\frac{x^2}{x^2-1},
\end{equation}
where $s_i$ is $\alpha$ or $1-\alpha$ depending on the type of excitation.

Let us introduce the notation
\begin{equation}
\epsilon_i=\int \frac{\rho_i(x)}{x^2}dx,
\end{equation}
and observe that
\begin{equation}
K_i=\int \frac{x^2-1}{x^2}\rho_i(x)dx=\int dx\, \sum_{k=1}^{K_i}\delta(x-x_{i,k}).
\end{equation}
It is now easy to see that the limit of the Bethe ansatz correctly reproduces the interaction terms that appear in the finite gap equations, as well as the winding term $\mathcal{M}$. In place of the residue of the quasi-momentum $\mathcal{E}$, in each equation there appear expressions involving the chain length $L$ and combinations of $\epsilon_i$ and $K_i$. If we denote such contributions by~$\mathcal{E}_i$, we find
\begin{align}
\mathcal{E}_1=\mathcal{E}_{\bar{1}}&=L+2  (\gamma +1) \epsilon _1-  \epsilon _2+
   (2   \Gamma +1 )\epsilon _3 +  (2 \gamma +1) \epsilon _{\bar{1}}+  \epsilon _{\bar{2}}+2  \Gamma  \epsilon _{\bar{3}}\nonumber\\
   &\ +  (   \gamma +1/2) K_1+  \Gamma  K_3+(\gamma +1/2 ) K_{\bar{1}} +\Gamma  K_{\bar{3}},\\
   \mathcal{E}_3=\mathcal{E}_{\bar{3}}&=L+ (2 \Gamma +1) \epsilon _1-\epsilon _2+2 (\gamma +1) \epsilon _3+2 \Gamma  \epsilon
   _{\bar{1}} +\epsilon _{\bar{2}}+(2 \gamma +1) \epsilon _{\bar{3}}\nonumber\\
   &\ +\Gamma  K_1+ ( \gamma +1/2) K_3 + \Gamma  K_{\bar{1}}+ ( \gamma +1/2)K_{\bar{3}}.
\end{align}
From the finite gap construction we expect  
\begin{equation}
\mathcal{E}_1=\mathcal{E}_{\bar{1}}=\mathcal{E}_3=\mathcal{E}_{\bar{3}}\equiv\mathcal{E},
\end{equation}
which is indeed possible if
\begin{equation}
\label{eq:fgcond}
\Gamma=\gamma+\frac{1}{2}.
\end{equation}
In this case the relation between $\mathcal{E}$ and $L$ is
\begin{equation}
\mathcal{E}=L+2(1+\gamma)(\epsilon_1+\epsilon_3)-\epsilon_2+(1+2\gamma)(\epsilon_{\bar{1}}+\epsilon_{\bar{3}})+\epsilon_{\bar{2}}+\big(\gamma+\frac{1}{2}\big)\left(K_{1}+K_{3}+K_{\bar{1}}+K_{\bar{3}}\right).
\end{equation}


\subsection{Comparison with the near-BMN limit}
We now consider the near-BMN expansion of our conjectured equations, where we again take $h\gg1$ but let the number of excitation be small, $K_i\ll L$~\cite{Berenstein:2002jq}. For this purpose, it is sufficient to recall that by imposing that momentum scales as $p=\mathsfit{p}/h$ where $\mathsfit{p}$ is finite when $h\gg1$, we find
\begin{equation}
x_i^\pm= \frac{s_i +\omega_{\mathsfit{p}_i}}{\mathsfit{p}_i}+O(h^{-1}),\quad \omega_{\mathsfit{p}_i}=\sqrt{s_i^2+\mathsfit{p}_i^2}.
\end{equation}

The expansion is straightforward, but it is interesting to observe the result in the simple case where only excitations of type 1 and 3 are present, following~\cite{Rughoonauth:2012qd}. Recall that semiclassically we take the scalar factors to be given by~(\ref{eq:dressing-ansatz11}) and~(\ref{eq:dressing-ansatz13}). It is then easy to compute
\begin{align}
-ih\log \frac{x^+_{p_1}-x^-_{q_1}}{x^-_{p_1}-x^+_{q_1}}S_{11}(p_1,q_1) &\approx\frac{\mathsfit{p}_1-\mathsfit{q}_1}{2} -\frac{s_1}{2}\frac{(\mathsfit{p}_1+\mathsfit{q}_1)^2}{\mathsfit{p}_1\,\omega_{\mathsfit{q}_1}-\mathsfit{q}_1\,\omega_{\mathsfit{p}_1}}\nonumber
-(1+2\gamma)\!\left(\frac{\mathsfit{p}_1\,\omega_{\mathsfit{q}_1}}{2\,s_1}-\frac{\mathsfit{q}_1\,\omega_{\mathsfit{p}_1}}{2\,s_1}\right)\!,\\
-ih\log \frac{z^+_{p_3}-z^-_{q_3}}{z^-_{p_3}-z^+_{q_3}}S_{33}(p_3,q_3) &\approx\frac{\mathsfit{p}_3-\mathsfit{q}_3}{2} -\frac{s_3}{2}\frac{(\mathsfit{p}_3+\mathsfit{q}_3)^2}{\mathsfit{p}_3\,\omega_{\mathsfit{q}_3}-\mathsfit{q}_3\,\omega_{\mathsfit{p}_3}}\nonumber
-(1+2\gamma)\!\left(\frac{\mathsfit{p}_3\,\omega_{\mathsfit{q}_3}}{2\,s_3}-\frac{\mathsfit{q}_3\,\omega_{\mathsfit{p}_3}}{2\,s_3}\right)\!,\\
-ih\log S_{ij}(p_i,q_j) &\approx \frac{\mathsfit{p}_i-\mathsfit{q}_j}{2}-(1+2\Gamma)\left(\frac{\mathsfit{p}_i\,\omega_{\mathsfit{q}_j}}{2\,s_j}-\frac{\mathsfit{q}_j\,\omega_{\mathsfit{p}_i}}{2\,s_i}\right).
\end{align}
When taking the limit of the all-loop Bethe equations, the terms in the first two lines proportional to $(\mathsfit{p}_i+\mathsfit{q}_i)^2$ correctly reproduce the one-loop S-matrices~\cite{Rughoonauth:2012qd}, whereas the remaining terms can be interpreted as shifts in the spin-chain length $L$. The limit of the equation for particles of type 1 and 3 then reads
\begin{equation}
\begin{aligned}
e^{i p_{1,k}L} &= e^{i \mathsfit{p}_{1,k}(\frac{1}{2}K_1-\frac{1+2\gamma}{2\,s_1} E_1+\frac{1}{2}K_3-\frac{1+2\Gamma}{2\,s_3}E_3)} \prod_{j\neq k}{S_{11}^{\text{1-loop}}(\mathsfit{p}_{1,k},\mathsfit{q}_{1,j})} ,\\
e^{i p_{3,k}L} &= e^{i \mathsfit{p}_{3,k}(\frac{1}{2}K_3-\frac{1+2\gamma}{2\,s_3} E_3+\frac{1}{2}K_1-\frac{1+2\Gamma}{2\,s_1}E_1)} \prod_{j\neq k}{S_{33}^{\text{1-loop}}(\mathsfit{p}_{3,k},\mathsfit{q}_{3,j})} ,
\end{aligned}
\end{equation}
so that the spin-chain length must scale as $L\approx 2\pi h$ when $h\gg1$. Here we used the short-hand notation for the total energy of each type of excitation
\begin{equation}
E_j=\sum_{k=1}^{K_j}\omega_{\mathsfit{p}_{j,k}},
\end{equation}
and assumed a configuration satisfying
\begin{equation}
\exp\left({i \sum_{k=1}^{K_1}{p_{1,k}}}\right)=1=\exp\left({i \sum_{k=1}^{K_3}{p_{3,k}}}\right).
\end{equation}
This particular level matching condition is a simplifying assumption that allows us to compare with the explicit near-BMN calculation of \cite{Rughoonauth:2012qd}.
In order to reproduce the energy shifts computed there, we must recast the one-loop Bethe equation in such a way that the equations of modes with different masses decouple, so that
\begin{equation}
e^{i \mathsfit{p}_{1,k}\ell}= \prod_{j\neq k}{S_{11}^{\text{1-loop}}(\mathsfit{p}_{1,k},\mathsfit{q}_{1,j})},\quad\quad
e^{i \mathsfit{p}_{3,k}\ell}= \prod_{j\neq k}{S_{33}^{\text{1-loop}}(\mathsfit{p}_{3,k},\mathsfit{q}_{3,j})},
\end{equation}
where the one-loop S-matrix must be
\begin{equation}
-ih\log S_{ii}^{\text{1-loop}}(\mathsfit{p}_{i},\mathsfit{q}_{i})
\approx -\frac{s_i}{2}\frac{(\mathsfit{p}_i+\mathsfit{q}_i)^2}{\mathsfit{p}_i\,\omega_{\mathsfit{q}_i}-\mathsfit{q}_i\,\omega_{\mathsfit{p}_i}}.
\end{equation}
We can do this by fixing the relation between the string length $\ell$ and the spin-chain length $L$ to be
\begin{equation}
\ell=L-\frac{1}{2}(K_1+K_3)+\frac{1+2\gamma}{2\,s_1}E_1+\frac{1+2\gamma}{2\,s_3}E_3 ,
\end{equation}
provided that 
\begin{equation}
\label{eq:bmncond}
\Gamma=\gamma.
\end{equation}
Surprisingly, this is inconsistent with the condition we found from the finite-gap construction~(\ref{eq:fgcond}). In order to further investigate this mismatch, let us directly compare the two semiclassical pictures.

\subsection{Comparing the coset construction to near-BMN expansion}
To highlight the mismatch between the near-BMN limit and the finite gap equations it is sufficient to consider the simpler case where only excitations of type 1 and 3 are present. We can also overlook the contributions due to winding terms.
The finite gap equations then simply read
\begin{align}
  2\pi n_1 &= -\frac{2\,s_1\,x}{x^2-1}2\pi\mathcal{E}+4\,s_1\pint\frac{\rho_1(y)}{x-y}dy\\
  2\pi n_3 &= -\frac{2\,s_3\,x}{x^2-1}2\pi\mathcal{E}+4\,s_3\pint\frac{\rho_3(y)}{x-y}dy.
\end{align}
If we now assume that the densities emerge from 
\begin{equation}
  \rho_i(x)=\sum_{j=1}^{K_i} \frac{x^2}{x^2-1}\,\delta(x-x_{i,j}),\quad\quad \text{with $x_{i}\approx\frac{s_i+\omega_{\mathsfit{p}_i}}{\mathsfit{p}_i}$} ,
\end{equation}
we can expand the interaction kernel as
\begin{equation}
  4\,s_1\pint\frac{\rho_1(y)}{x-y}dy \approx
  \sum_{j\neq k}^{K_1} \frac{s_1\,(\mathsfit{p}_{1,k}+\mathsfit{p}_{1,j})^2}{2(\mathsfit{p}_{1,k}\omega_{\mathsfit{p}_{1,j}}-\mathsfit{p}_{1,j}\omega_{\mathsfit{p}_{1,k}})}+ \frac{1}{2\,s_1}\mathsfit{p}_{1,k}E_1 ,
\end{equation}
 so that
\begin{equation}
2\pi n_1 = -\mathsfit{p}_{1,k}\left(2\pi\mathcal{E}+\frac{1}{2\,s_1}E_1\right)+\sum_{j\neq k}^{K_1}\log {S_{11}^{\text{1-loop}}(\mathsfit{p}_{1,k},\mathsfit{p}_{1,j})},
\end{equation}
and likewise
\begin{equation}
  2\pi n_3 = -\mathsfit{p}_{3,k}\left(2\pi\mathcal{E}+\frac{1}{2\,s_3}E_3\right)+\sum_{j\neq k}^{K_3}\log {S_{33}^{\text{1-loop}}(\mathsfit{p}_{3,k},\mathsfit{p}_{3,j})}.
\end{equation}
The coefficients of the terms that are linear in the momenta should be identified with  the string length $\ell$. However, the two coefficients are not equal. Therefore, it appears impossible to find a unique relation between~$\mathcal{E}$ and~$\ell$.
It would seem that this mismatch does not depend on the details of our all-loop construction or on the ansatz for the scalar factors that we conjecture, but emerges already at the level of the semiclassical constructions.

It is worth remarking that the S-matrix elements can indeed shift by terms of the form $p_i E_j$ when modifying the choice of the light-cone gauge, even if the spectrum should be gauge invariant. Therefore a possible way to investigate the mismatch could be to repeat the calculation of~\cite{Rughoonauth:2012qd} by allowing for a more general light-cone gauge such as the $a$-gauge of~\cite{Arutyunov:2006gs} in place of the uniform light-cone gauge~\cite{Arutyunov:2005hd}.

Another possibility is that the algebraic curve does not quite reproduce the original string theory. To some extent, this was always known as the algebraic curve construction overimposes the Virasoro conditions, requiring the worldsheet momentum to vanish separately on the two three-spheres, which results in the elimination of a physical massless mode from the spectrum. Perhaps, in addition to that, the requirement that the residue of the quasimomentum $\mathcal{E}$ is the same on cuts of type $1$ and $3$ is also an artifact of the construction, and we should really allow $\mathcal{E}_1\neq \mathcal{E}_3$. Unfortunately we are at the moment unable to determine which possibility is correct.

\section{Discussion and outlook}
Building on~\cite{Borsato:2012ud}, we have written down a set of all-loop Bethe equations for the $\alg{d}(2,1;\alpha)^2$ symmetric alternating spin-chain. These should reproduce the asymptotic spectrum of massive excitations of strings on $\AdS_3\times \Sphere^3\times \Sphere^3\times \Sphere^1$. However, before the quantum behavior of the massless modes of $\AdS_3\times \Sphere^3\times \Sphere^3\times \Sphere^1$ is understood we cannot be sure that our Bethe ansatz truly describes a consistent sector of the string theory, and indeed understanding the massless modes is one of the most important challenges for integrability in $\AdS_3/\CFT_2$.

Our Bethe ansatz is naturally written in a grading of $\alg{d}(2,1;\alpha)^2$ different from and seemingly inequivalent to the one conjectured earlier~\cite{Babichenko:2009dk,OhlssonSax:2011ms}.  The equations involve four undetermined scalar factors that play the role of a dressing phase. In contrast with what happens in $\AdS_5/\CFT_4$ and $\AdS_4/\CFT_3$, it appears that crossing symmetry imposes that these factors differ from the BES phase~\cite{Beisert:2006ez}, that seems to be consistent with the recent findings of~\cite{Abbott:2012dd,Beccaria:2012kb}. Here we have investigated their form only in the large $h$ limit (\ie, semi-classical strings), and established that they must also non-trivially couple nodes of different mass -- another new feature with respect to~\cite{Babichenko:2009dk,OhlssonSax:2011ms}.

It seems that in the semiclassical limit all scalar factors can be expressed in terms of an AFS-like phase~\cite{Arutyunov:2004vx}. This is also compatible with~\cite{Abbott:2012dd}, where such a form is necessary to reproduce the L\"uscher corrections computed from the finite gap equations. However, when we tried to fix the semiclassical scalar factors we apparently encountered a contradiction between the finite gap construction of~\cite{Babichenko:2009dk,Zarembo:2010yz} and the near-BMN expansion of~\cite{Rughoonauth:2012qd}. At the moment it is not clear how this mismatch will be resolved. This is a further motivation to solve the crossing equations at all-loop, which may be a non-trivial task. Besides the analytical complexity of the problem, this may require some physical input on the bound-state spectrum of the theory. As remarked in~\cite{Borsato:2012ud}, representation theory of centrally extended $\algSU(1|1)^2$ allows for bound states of particles of different mass. If such modes are present one should allow for suitable poles in the corresponding scalar factors.

Solving the crossing equations for the $\alg{d}(2,1;\alpha)^2$ spin-chain would allow to address several physical issues. A significant one is the $\alpha\to0$ limit of the theory, where one sphere blows up and two modes become massless. Upon compactification, this would give an $\AdS_3\times \Sphere^3\times \Torus^4$ background, that also appears to be integrable~\cite{Babichenko:2009dk,Zarembo:2010yz,Beccaria:2012kb}. Understanding how these massless modes decouple may shed light on how to incorporate the $\AdS_3\times \Sphere^3\times \Sphere^3\times \Sphere^1$ modes that the are not captured by the coset construction. A first effort to understand this was taken in~\cite{Sax:2012jv} and it would be interesting to continue the investigation at all-loop.
It would also be interesting to investigate whether quantum integrability can be extended to $\AdS_3$ backgrounds containing NS-NS flux~\cite{Cagnazzo:2012se}.
We hope to return to some of these questions in the future.

\section*{Acknowledgments}

We would like to thank M.~Abbott, G.~Arutyunov, M.~de~Leeuw, P.~Sundin, S.~van Tongeren and K.~Zarembo for valuable discussions and useful comments.
The authors acknowledge support by the Netherlands Organization for Scientific Research (NWO) under the VICI grant 680-47-602.
The work by the authors is also part of the ERC Advanced grant research programme No. 246974, ``Supersymmetry: a window to non-perturbative physics''.


\bibliographystyle{oos}
\bibliography{refs}

\end{document}